%% file: main.tex
\documentclass{article} 
\usepackage[dvipsnames,table,xcdraw]{xcolor}
\usepackage{preprint,times}

\input{math_commands.tex}

\usepackage{hyperref}
\usepackage{url}
\usepackage{amsmath,amssymb,amsthm,mathtools}
\usepackage{booktabs,multirow,array}
\usepackage{longtable}
\usepackage{graphicx}
\usepackage{enumitem}
\setlist{nosep,wide=0pt,before=\vspace{-\parskip}}
\usepackage{float}
\newcolumntype{?}{!{\color{black!35}\vrule width 0.4pt}}
\renewcommand{\arraystretch}{1.18}

\newcommand{\Hs}{\mathbf{H}^\star}
\newcommand{\Hh}{\hat{\mathbf{H}}}
\newcommand{\dH}{\delta\mathbf{H}}
\newcommand{\Sm}{\mathbf{S}}
\newcommand{\Cs}{\mathbf{C}^\star}
\newcommand{\Ch}{\hat{\mathbf{C}}}
\newcommand{\epsv}{\boldsymbol{\varepsilon}}
\newcommand{\Dl}{\boldsymbol{\Delta}}
\newcommand{\Pm}{\mathbf{P}}
\newcommand{\occ}{\mathcal{O}}
\newcommand{\vir}{\mathcal{V}}
\newcommand{\Lelem}{\mathcal{L}_{\mathrm{elem}}}
\newcommand{\LWA}{\mathcal{L}_{\mathrm{WA}}}
\newcommand{\Lrho}{\mathcal{L}_{\rho}}
\newcommand{\Lldos}{\mathcal{L}_{\mathrm{LDOS}}}
\newcommand{\Lcum}{\mathcal{L}_{\mathrm{cum}}}
\newcommand{\Nre}{N}
\newcommand{\uHa}{$\mu$Ha}

\colorlet{rowbase}{gray!13}
\colorlet{rowabl}{violet!14}
\colorlet{rowours}{blue!13}

\DeclareMathOperator{\diag}{diag}
\DeclareMathOperator{\erf}{erf}
\DeclareMathOperator{\erfc}{erfc}

\DeclareMathOperator{\spn}{span}
\DeclareMathOperator{\sgn}{sign}

\theoremstyle{plain}
\newtheorem{theorem}{Theorem}[section]
\newtheorem{proposition}[theorem]{Proposition}
\newtheorem{lemma}[theorem]{Lemma}

\theoremstyle{definition}
\newtheorem{definition}[theorem]{Definition}
\theoremstyle{remark}
\newtheorem{remark}[theorem]{Remark}

\title{\texorpdfstring{\raggedright}{}Where the Error Lives: \\ Spectral Representations of \\ Learned Kohn--Sham Hamiltonians}

\author{Fang Sun$^{*}$, Hekai Cheng$^{*}$, Ruoyan Li, Yuanzhou Chen, \\
\textbf{Wei Wang, Jason Cong \& Yizhou Sun} \\
Computer Science Department\\
University of California, Los Angeles}

\begin{document}

\maketitle
{\renewcommand{\thefootnote}{\fnsymbol{footnote}}\footnotetext[1]{Equal contribution.}}

\begin{abstract}
Kohn--Sham density functional theory (DFT) is the standard method for computing the electronic structure of molecules and materials, but its self-consistent-field (SCF) iterations make it expensive for large systems and high-throughput screening. Machine-learned Hamiltonian models offer a shortcut: an equivariant network predicts the Kohn--Sham Hamiltonian matrix in one forward pass, and diagonalizing the prediction yields orbital energies, the HOMO--LUMO gap, molecular orbitals and the electron density. These models are trained to match the reference matrix entry by entry, yet a small entry-wise error does not guarantee accurate eigenvalues or eigenvectors. Two physics-informed alternatives go partway: a recently proposed wavefunction-alignment loss adds a spectral penalty without diagonalizing the prediction, and matching the electron density of the diagonalized prediction supervises it directly. Neither is complete: by representing the prediction error in the eigenbasis of the reference Hamiltonian, we show that each sees only part of the error, and that shifting the predicted orbital energies leaves the electron density exactly unchanged. The same analysis identifies a complete training signal: the \emph{cumulative local density of states} (cumulative LDOS), the number of electrons at each point in space with energy below a given threshold. The cumulative LDOS is a single physical observable that contains the electron density as a special case, responds to every component of the error, determines the Hamiltonian uniquely, and yields gradients that do not vanish for large orbital-energy errors. Supervising through it, a method we call \emph{FermiScope}, requires no change to the network architecture. Fine-tuning three Hamiltonian networks (QHNet, SPHNet and WANet) on QH9 in this way reduces the HOMO--LUMO gap error by 49--79\%, 2.0--2.6 times more than the wavefunction-alignment loss, without increasing the element-wise error by more than 0.3\%. The gains carry over to out-of-distribution and molecular-dynamics splits (36--77\%), where the wavefunction-alignment loss helps less. These results show that how a predicted Hamiltonian is represented during training, not only the architecture that produces it, decides whether it is useful in practice.
\end{abstract}

\section{Introduction}
\label{sec:intro}

Kohn--Sham density functional theory (DFT) \citep{hohenberg1964,kohn1965}, recognized by the 1998 Nobel Prize in Chemistry \citep{kohn1999nobel}, is the most widely used method for computing the electronic structure of molecules and materials. It summarizes the electrons of a molecule in a single matrix, the Kohn--Sham Hamiltonian, whose generalized eigendecomposition gives the orbital energies, the energies of the highest occupied and lowest unoccupied molecular orbitals (HOMO and LUMO) and their gap, the molecular orbitals, and the electron density. These quantities determine chemical reactivity, optical and electronic behaviour and, through the total energy, molecular stability. Obtaining the Hamiltonian, however, requires self-consistent-field (SCF) iterations \citep{szabo1996modern}. Each iteration builds the Hamiltonian from the current occupied orbitals ($O(n^4)$ in the number of basis functions $n$ for hybrid functionals) and diagonalizes it to obtain new ones ($O(n^3)$). A typical calculation repeats this $T$ times, until the Hamiltonian and the orbitals it is built from agree ($T\approx9$ for the QH9 molecules studied here; Appendix~\ref{app:ks}). This loop limits DFT on large systems and in high-throughput screening.

\begin{figure}[t]
\centering
\includegraphics[width=\textwidth]{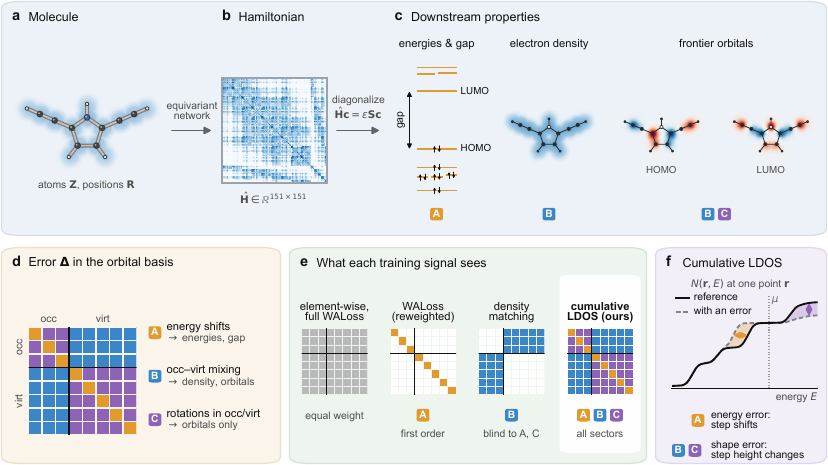}
\vspace{-1.5em}
\caption{\textbf{Overview.} \textbf{(a--c)} A Hamiltonian network maps a molecule, QH9 $\#23470$ (C$_8$H$_5$N), to its Kohn--Sham Hamiltonian $\hat{\mathbf H}$, whose eigendecomposition gives the orbital energies and HOMO--LUMO gap, the electron density and the molecular orbitals. All visualizations in panels~(a--c) use reference DFT data for this molecule. \textbf{(d)} In the reference orbital basis, the prediction error splits into three sectors: orbital-energy shifts (\textcolor{YellowOrange}{\textbf{A}}, yellow-orange diagonal), occupied--virtual mixing (\textcolor{RoyalBlue}{\textbf{B}}, blue off-diagonal block) and rotations within the occupied or the virtual space (\textcolor{violet}{\textbf{C}}, purple off-diagonal blocks). The tags in panel~(c) mark the sectors each downstream property depends on. \textbf{(e)} The sectors each training signal sees. WALoss is from \citet{li2025wanet}; density matching, \Eqref{eq:lrho}, is the $E=\mu$ slice of our loss. \textbf{(f)} At a point $\mathbf r$, the cumulative LDOS $N(\mathbf r,E)$ is a staircase with one step per orbital. An energy error shifts a step (\textcolor{YellowOrange}{\textbf{A}}), and a shape error changes step heights (\textcolor{RoyalBlue}{\textbf{B}}, \textcolor{violet}{\textbf{C}}). The shape error shown changes the LUMO step but not the slice at $\mu$, which is the density, so it is of type \textcolor{violet}{\textbf{C}} and invisible to density matching; Figure~\ref{fig:fingerprints} shows all three types.}
\label{fig:teaser}
\vspace{-1em}
\end{figure}

Machine-learned Hamiltonian models remove this bottleneck. An equivariant neural network predicts the Hamiltonian directly from the atomic numbers and positions \citep{schutt2019schnorb,unke2021phisnet,li2022deeph,yu2023qhnet,li2025wanet,luo2025sphnet}, replacing all $T$ iterations by one forward pass. The predicted Hamiltonian is used in two ways: it can be diagonalized once to obtain orbital energies, orbitals and the density, or it can initialize a conventional DFT calculation to save SCF iterations \citep{yu2023qh9,li2025wanet}. We focus on the first setting, where no SCF iteration is left to correct the prediction and the quality of the loss function directly determines the quality of the result. On the QH9 benchmark \citep{yu2023qh9}, the entry-wise error of the predicted Hamiltonian has more than halved since the benchmark was introduced (Table~\ref{tab:main}), but this has not always translated into better eigenvalues.

The entry-wise error does not translate directly into accurate eigenvalues and eigenvectors. For example, continuing to train QHNet with its own element-wise loss lowers its matrix error but raises its HOMO--LUMO gap error, consistently on all four QH9 splits and by 53\% on the standard split (Appendix~\ref{app:honly}). \citet{li2025wanet} observed the same decoupling from the other side: on QH9, their alignment loss increased the matrix error while improving the orbital energies, and on larger molecules, matrices with tiny relative errors produced unusable total energies. The reason is that these models are trained with an \emph{element-wise} loss, the mean absolute or squared error over matrix entries, which treats every entry alike, whereas eigenvalues and eigenvectors respond very unevenly to an error. Classical matrix perturbation theory makes this precise \citep{kato1995perturbation,davis1970rotation,stewart1990perturbation}: to first order, an eigenvalue moves only with the corresponding diagonal entry of the error expressed in the eigenbasis, and the eigenvectors rotate by the off-diagonal entries divided by eigenvalue differences. On QH9, an error that couples the HOMO and the LUMO, about 0.25\,Ha apart, therefore changes the LUMO about 75 times more than an equally large error that couples it to a core orbital some 19\,Ha lower.

To make this precise, we represent the prediction error in the eigenbasis of the reference Hamiltonian, where it separates into three interpretable blocks, which we call \emph{sectors} (Figure~\ref{fig:teaser}d): diagonal entries, which shift orbital energies; entries corresponding to an occupied and an unoccupied orbital, which mix the two and change the electron density; and entries corresponding to two occupied or two unoccupied orbitals, which only rotate orbitals within those spaces. This view explains why existing physics-informed losses help only partly. The wavefunction-alignment loss (WALoss) of \citet{li2025wanet} evaluates the predicted Hamiltonian in the reference orbital basis. In its full form it weights every entry of the error uniformly, like the element-wise loss, and therefore does not concentrate on the errors that matter; in the diagonal-only form proposed by \citet{li2025wanet}, it penalizes only the orbital energies, and only to first order. A natural alternative is to diagonalize the prediction and match its electron density, which is exact and automatically focuses on the frontier orbitals, those closest to the HOMO--LUMO gap. We prove, however, that energy shifts and rotations within the occupied or the unoccupied space leave the density exactly unchanged, so a loss that depends only on the density does not respond to them at all; we say that it is \emph{blind} to them. These errors span about 69\% of the degrees of freedom of the error on QH9 and include every orbital-energy error.

The same representation tells us what a complete training signal must do: see every sector, weight each by its effect on the properties of interest, and use the exact eigenpairs of the prediction. We find such a signal in a physical observable, the \emph{local density of states} (LDOS), which resolves the electron density by energy and is what a scanning tunnelling microscope measures \citep{tersoff1985stm}. Its cumulative form, $N(\mathbf{r},E)$, the number of electrons at point $\mathbf{r}$ with energy below $E$, represents the Hamiltonian in the language of the density without discarding the orbital energies (Figure~\ref{fig:teaser}f). We prove four properties. The density is its slice at the Fermi level, the energy midway between the HOMO and the LUMO. It responds to every sector. It determines the Hamiltonian uniquely. And unlike a peak-shaped LDOS, it keeps pulling an orbital energy toward its target however far inside the energy window it starts. We turn $N$ into a training loss, the \emph{cumulative LDOS loss} $\Lcum$: the mean squared difference between $N$ computed from the predicted and from the reference Hamiltonian, over sampled points $\mathbf{r}$ and energies $E$. A network is fine-tuned on $\Lelem+w\,\Lcum$, where $\Lelem$ is its usual element-wise loss and $w$ a weight, with no change to its architecture. We call this method \emph{FermiScope}, as it scans $N$ across a window of energies around the Fermi level.

Across three architectures and four QH9 splits, this loss improves every eigen-derived metric in all twelve settings and is the best method in each (Tables~\ref{tab:main} and~\ref{tab:splits}). On the standard split, it lowers the HOMO--LUMO gap error by 49--79\%, 2.0--2.6 times more than WALoss, while the element-wise error rises by at most 0.3\% and on QHNet even falls by 3.5\%. Under distribution shift, it still lowers the gap error by 36--77\%, whereas WALoss makes it worse in three of the nine shifted settings, all on QHNet.

Our contributions are as follows.
\begin{itemize}
\item \textbf{Problem.} We show that the mismatch between how Hamiltonian networks are trained, entry by entry, and how they are used, through eigenpairs, is a question of \emph{where} the error lies. In the reference orbital basis, the error separates into sectors that each property and each training signal read selectively, and the electron density, the most natural physical target, is blind to every orbital-energy error (Sec.~\ref{sec:analysis}).
\item \textbf{Solution.} We identify the cumulative LDOS as a representation of the Hamiltonian that contains the density, sees every sector and determines the Hamiltonian uniquely, and we turn it into a training loss that any Hamiltonian network can use unchanged (Secs.~\ref{sec:loss} and~\ref{sec:props}).
\item \textbf{Results and implications.} The loss improves every eigen-derived metric in all twelve backbone--split settings without raising the element-wise error by more than 0.3\% (Sec.~\ref{sec:exp}). How a predicted Hamiltonian is represented during training, and not only the architecture that produces it, decides whether it is useful.
\end{itemize}

\section{Background}
\label{sec:background}
\paragraph{Notation.}
We follow standard quantum-chemistry notation \citep{szabo1996modern}; Appendix~\ref{app:ks} explains the terms we use and illustrates the SCF procedure. Bold upright symbols denote matrices and vectors. A star marks quantities of the reference DFT calculation and a hat quantities computed from a network's prediction; unadorned symbols ($\varepsilon$, $\mathbf{C}$, $\psi$, $\rho$, $N$) denote the same maps applied to a generic Hamiltonian. A hat thus means \emph{predicted}, as in statistics.

\paragraph{Kohn--Sham Hamiltonians.}
Consider a molecule of $N_{\mathrm{at}}$ atoms with atomic numbers $\mathbf{Z}\in\mathbb{Z}_{+}^{N_{\mathrm{at}}}$, which identify the element of each atom, and positions $\mathbf{R}\in\mathbb{R}^{3N_{\mathrm{at}}}$. Its electrons are described in a basis of $n$ real-valued, atom-centred basis functions $\{\phi_\mu:\mathbb{R}^3\to\mathbb{R}\}_{\mu=1}^{n}$, also called atomic orbitals; for the def2-SVP basis used here, they are Gaussian-type functions. The basis functions are linearly independent but not orthogonal. Their inner products form the overlap matrix $\Sm$, with entries $S_{\mu\nu}=\int\phi_\mu(\mathbf{r})\phi_\nu(\mathbf{r})\,d\mathbf{r}$; it depends only on the geometry and is positive definite, written $\Sm\succ0$. In Kohn--Sham DFT \citep{hohenberg1964,kohn1965}, the electronic structure is represented by a second symmetric $n\times n$ matrix, the Kohn--Sham Hamiltonian $\Hs$, with entries $H^\star_{\mu\nu}=\langle\phi_\mu|\mathcal{F}[\Pm^\star]|\phi_\nu\rangle$, where $\mathcal{F}[\Pm]$ is the Kohn--Sham operator built from the occupied orbitals through their density matrix $\Pm$ (\Eqref{eq:density}) and $\Pm^\star$ is the converged, self-consistent one. The molecular orbitals of the reference are linear combinations of the basis functions, $\psi_i(\mathbf{r})=\sum_\mu C^\star_{\mu i}\phi_\mu(\mathbf{r})$ (we omit their star), whose coefficients and energies solve the generalized eigenvalue problem
\begin{equation}
\Hs\Cs=\Sm\Cs\diag(\epsv^\star),\qquad \Cs{}^\top\Sm\Cs=\mathbf{I}.
\label{eq:gep}
\end{equation}
The second condition states that, unlike the basis functions, the molecular orbitals are orthonormal, $\int\psi_i(\mathbf{r})\psi_j(\mathbf{r})\,d\mathbf{r}=\delta_{ij}$. Column $\mathbf{c}^\star_i$ of $\Cs$ holds the coefficients of orbital $\psi_i$, and $\varepsilon^\star_i$ is its energy. A closed-shell molecule with $N_e$ electrons occupies the $n_o=N_e/2$ lowest orbitals (the set $\occ$); the remaining $n_v=n-n_o$ orbitals are unoccupied, or virtual (the set $\vir$). The HOMO and LUMO energies are $\varepsilon^\star_{n_o}$ and $\varepsilon^\star_{n_o+1}$, and the electron density is
\begin{equation}
\rho^\star(\mathbf{r})=2\sum_{i\in\occ}|\psi_i(\mathbf{r})|^2=\boldsymbol\phi(\mathbf r)^\top\Pm^\star\boldsymbol\phi(\mathbf r),\qquad \Pm^\star=2\,\Cs_{\occ}\Cs_{\occ}{}^\top,
\label{eq:density}
\end{equation}
where $\boldsymbol\phi=(\phi_1,\dots,\phi_n)^\top$ and $\Cs_{\occ}$ holds the occupied columns of $\Cs$. The density matrix $\Pm^\star$ is twice the projector onto the occupied subspace, the span of the $n_o$ lowest orbitals.

Two maps connect the Hamiltonian and the density matrix. Diagonalizing a Hamiltonian and filling its $n_o$ lowest orbitals gives a density matrix, $\mathbf H\mapsto\Pm[\mathbf H]$, and from it the density by \Eqref{eq:density}. Conversely, evaluating the Kohn--Sham operator in the basis builds a Hamiltonian from a density matrix, $\Pm\mapsto\mathbf H[\Pm]$; for a hybrid functional such as B3LYP, the exact-exchange term needs all of $\Pm$, not only the density. DFT alternates the two maps in the self-consistent-field (SCF) iteration,
\begin{equation}
\mathbf H^{(t)}\;\xrightarrow{\ \text{diagonalize}\ }\;\big(\epsv^{(t)},\mathbf C^{(t)}\big)\;\xrightarrow{\ \text{fill } n_o\ }\;\Pm^{(t)}\;\xrightarrow{\ \text{build}\ }\;\mathbf H^{(t+1)}=\mathbf H\big[\Pm^{(t)}\big],
\label{eq:scf}
\end{equation}
from an initial guess until $\Pm$ stops changing. The reference is the fixed point of this loop, $\Hs=\mathbf H[\Pm^\star]$ with $\Pm^\star=\Pm[\Hs]$. The two maps are not mirror images. The build gives one Hamiltonian for each density matrix, but many Hamiltonians share the same density matrix, because $\Pm$ records only which subspace is occupied, not the orbital energies or how the subspace is split into individual orbitals. The build costs $O(n^4)$ for hybrid functionals, and the $T$ iterations dominate the cost of the calculation (Figure~\ref{fig:appscf} in Appendix~\ref{app:ks}).

\paragraph{Learning Hamiltonians.}
A Hamiltonian network predicts $\Hh=f_\theta(\mathbf{Z},\mathbf{R})$ and is built to transform correctly under rotations of the molecule; $\Sm$ is computed exactly from the geometry. Existing models are trained with the element-wise loss $\Lelem=\|\Hh-\Hs\|$, the mean absolute or squared error over matrix entries \citep{yu2023qhnet,li2025wanet,luo2025sphnet}. At inference, $\Hh$ replaces $\Hs$ in \Eqref{eq:gep}, and every downstream quantity follows from its eigenpairs $(\hat\epsv,\Ch)$. Following QH9 \citep{yu2023qh9}, models are evaluated by the element-wise error of $\Hh$ together with the errors of the orbital energies and the similarity of the orbital coefficients (Sec.~\ref{sec:setup}).

\paragraph{Physics-informed losses.}
The wavefunction-alignment loss (WALoss) of \citet{li2025wanet} adds a spectral signal without diagonalizing the prediction. It precomputes the reference orbitals $\Cs$ and evaluates the prediction in their basis: since $\Cs$ diagonalizes $\Hs$, the matrix $\Cs{}^\top\Hh\Cs$ should equal $\diag(\epsv^\star)$. The full form penalizes every entry of the deviation alike, $\LWA=\ell\big(\Cs{}^\top\Hh\Cs-\diag(\epsv^\star)\big)$, where $\ell$ is the mean absolute or squared error over the entries (Appendix~\ref{app:dose}); we use this form for all backbones. \citet{li2025wanet} also propose a diagonal-only form that keeps only the diagonal, $\sum_i w_i\big(\mathbf{c}_i^{\star\top}\Hh\,\mathbf{c}^\star_i-\varepsilon^\star_i\big)^2$, with large weights on the occupied orbitals and the LUMO; it appears as \emph{reweighted} WALoss in Figure~\ref{fig:teaser}e. A second option is to diagonalize $\Hh$, form the predicted density $\rho[\Hh]$ by \Eqref{eq:density}, and penalize its deviation from $\rho^\star$ at a set $\mathcal{R}$ of sample points,
\begin{equation}
\Lrho(\Hh)=\Big\langle\Big(\frac{\rho[\Hh](\mathbf{r})-\rho^\star(\mathbf{r})}{\mathrm{div}(\mathbf{r})}\Big)^2\Big\rangle_{\mathbf{r}\in\mathcal{R}},
\label{eq:lrho}
\end{equation}
where $\mathrm{div}(\mathbf{r})\ge1$ is a geometry-only factor that down-weights the region near the nuclei, where the density is orders of magnitude larger \citep{ushenin2025lagnet}. We call \Eqref{eq:lrho} density matching; Sec.~\ref{sec:loss} shows that it is a special case of our loss.

\section{Spectral Representations of Hamiltonian Errors}
\label{sec:method}

We first represent the prediction error in the basis of reference orbitals and determine which part of it each eigen-property depends on (Sec.~\ref{sec:analysis}). This identifies what a complete training signal must see and leads to the cumulative LDOS (Sec.~\ref{sec:loss}), whose guarantees we establish in Sec.~\ref{sec:props}.

\subsection{Where the prediction error lives}
\label{sec:analysis}

Write the prediction as $\Hh=\Hs+\dH$ and express the error in the basis of reference orbitals,
\begin{equation}
\Dl:=\Cs{}^\top\dH\,\Cs,\qquad\text{so that}\qquad \Cs{}^\top\Hh\,\Cs=\diag(\epsv^\star)+\Dl .
\label{eq:Delta}
\end{equation}
Entry $\Delta_{ij}$ measures how strongly the error couples reference orbitals $i$ and $j$. We partition the symmetric matrix $\Dl$ into three \emph{sectors} (Figure~\ref{fig:teaser}d): (\textcolor{YellowOrange}{\textbf{A}}) the diagonal entries $d_i:=\Delta_{ii}$; (\textcolor{RoyalBlue}{\textbf{B}}) the occupied--virtual block $\Dl_{\occ\vir}$; and (\textcolor{violet}{\textbf{C}}) the off-diagonal parts of the occupied--occupied and virtual--virtual blocks, $\Dl_{\occ\occ}$ and $\Dl_{\vir\vir}$. This partition is exact: it sorts the entries of $\Dl$ by position alone, according to whether an entry lies on the diagonal and whether the orbitals it couples are occupied or virtual. The full WALoss averages the entries of $\Dl$ without weights (Lemma~\ref{lem:whitening}), so it treats all sectors alike.

Classical matrix perturbation theory \citep{kato1995perturbation,stewart1990perturbation} then shows what each sector does to the downstream properties.

\begin{proposition}[What each property depends on]
\label{prop:sectors}
Let the reference orbital energies be non-degenerate, and expand the eigenvalues and eigenvectors of $\Hh$ as functions of $\Dl$ around the reference, $\Dl=\mathbf 0$. To first order:
\begin{enumerate}[label=(\alph*)]
\item \emph{Sector~\textcolor{YellowOrange}{A}}: orbital energies depend only on the diagonal: $\hat\varepsilon_i-\varepsilon^\star_i=d_i$;
\item \emph{Sectors~\textcolor{RoyalBlue}{B} and \textcolor{violet}{C}}: orbital shapes depend on the off-diagonal entries, weighted by inverse energy spacings: $\hat{\mathbf{c}}_i-\mathbf{c}^\star_i=\sum_{j\ne i}\frac{\Delta_{ji}}{\varepsilon^\star_i-\varepsilon^\star_j}\,\mathbf{c}^\star_j$;
\item \emph{Sector~\textcolor{RoyalBlue}{B} only}: the density depends only on the occupied--virtual block, with the largest weights on orbital pairs near the HOMO--LUMO gap:
\begin{equation}
\delta\rho(\mathbf{r})=4\sum_{i\in\occ,\,a\in\vir}\frac{\Delta_{ai}\,\psi_i(\mathbf{r})\psi_a(\mathbf{r})}{\varepsilon^\star_i-\varepsilon^\star_a}.
\label{eq:drho}
\end{equation}
\end{enumerate}
\end{proposition}

Appendix~\ref{app:proofs} gives the proof and the second-order corrections. Item (c) holds because the occupied--occupied contributions cancel in pairs: rotating occupied orbitals among themselves leaves their total density unchanged. This cancellation is exact, not only first-order. Items (a) and (b) need non-degenerate orbital energies, but (c) needs only a nonzero HOMO--LUMO gap, because the density matrix is the spectral projector onto the occupied subspace, which depends analytically on the Hamiltonian while the gap is open \citep{kato1995perturbation}. This matters on QH9, where 12\% of the molecules have two orbital energies closer than 0.1\,mHa. Remark~\ref{rem:sectors} in Appendix~\ref{app:proofs} relates the sectors to the reported metrics.

\begin{proposition}[The density is blind to sectors \textcolor{YellowOrange}{A} and \textcolor{violet}{C}]
\label{prop:blind}
If $\Dl_{\occ\vir}=0$ and no eigenvalue of $\Hh$ crosses the HOMO--LUMO gap, then $\rho[\Hh]=\rho^\star$ exactly. Any loss that depends on $\Hh$ only through its density therefore vanishes on a subspace of dimension $\tfrac{n_o(n_o+1)}{2}+\tfrac{n_v(n_v+1)}{2}$ out of $\tfrac{n(n+1)}{2}$.
\end{proposition}

The reason is the map $\mathbf H\mapsto\Pm[\mathbf H]$ of Section~\ref{sec:background}, which keeps only the occupied subspace. If $\Dl_{\occ\vir}=0$, the prediction does not mix occupied with virtual reference orbitals, and if no level crosses the gap, its occupied subspace is that of the reference. Shifting energies (\textcolor{YellowOrange}{\textbf{A}}) or rotating orbitals within the occupied or the virtual space (\textcolor{violet}{\textbf{C}}) leaves this subspace, and with it $\Pm$ and $\rho$, in place.

On a uniform sample of 4{,}000 QH9-stable molecules, with median $n=171$ and $n_o=33$, this blind subspace holds a median of $69.0\%$ of the degrees of freedom of $\Dl$, that is, of its $n(n+1)/2$ independent entries, and 66--71\% between the 5th and 95th percentiles (Table~\ref{tab:datastats}). It contains every orbital-energy error, including the HOMO--LUMO gap error. The fraction counts directions, not error size: on physically realized Hamiltonian changes, the blind sectors carry 16--30\% of $\|\Dl\|_F^2$, more at larger displacements (Appendix~\ref{app:realresidual}). On a real molecule, shifting the HOMO or LUMO energy by 2\,mHa changes the density loss by less than $10^{-30}$, whereas rotating the HOMO and LUMO into each other changes it by $3\times10^{-7}$ (Figure~\ref{fig:sector}a in Appendix~\ref{app:verify}).

\begin{remark}[Why density matching still improves the gap]
\label{rem:blind}
Proposition~\ref{prop:blind} is a statement about directions, not about what a trained network does. A network produces all sectors from the same atomic-orbital blocks, so correcting the frontier couplings of sector~\textcolor{RoyalBlue}{B}, which the density weights by $1/(\varepsilon^\star_a-\varepsilon^\star_i)$, also changes the frontier orbital energies. In practice it improves them, and density matching alone lowers the gap error substantially (Sec.~\ref{sec:ablation}). The density loss is, however, exactly flat along any change of the predicted orbital energies that leaves the predicted orbitals fixed and moves no level across the gap, so it cannot correct the energy error that remains. That error lies in sector~\textcolor{YellowOrange}{A}, which is where energy resolution adds most (Sec.~\ref{sec:ablation}).
\end{remark}

\paragraph{From the density back to the Hamiltonian.}
SCF fixes the Hamiltonian from the density matrix, $\Hs=\mathbf H[\Pm^\star]$, but only because the reference is self-consistent: it is built from its own density matrix. A predicted $\Hh$ is not, and the step from $\Hh$ to its density matrix loses information. Every Hamiltonian that differs from $\Hs$ only in sectors~\textcolor{YellowOrange}{\textbf{A}} and \textcolor{violet}{\textbf{C}}, and moves no level across the gap, has the same $\Pm$ and hence the same $\rho$. Matching the density, or even the density matrix, therefore cannot single out $\Hs$. This is enough when the prediction only initializes SCF, whose first build turns the right $\Pm[\Hh]$ into $\Hs$ (Appendix~\ref{app:lossimpl}), but not when the prediction is diagonalized once and used directly, the setting we focus on.

\paragraph{What existing losses see.}
Figure~\ref{fig:teaser}e applies these results to the losses of Sec.~\ref{sec:background}, and Table~\ref{tab:coverage} in Appendix~\ref{app:proofs} lists how each loss weights the sectors. The element-wise loss and the full WALoss penalize every sector, but uniformly: a small coupling between two frontier orbitals, which shifts the gap at second order with weight $1/\text{gap}$ (Appendix~\ref{app:proofs}), costs as much as an equally large coupling between two deep core orbitals, which barely affects the properties of interest. The diagonal-only WALoss responds only to the diagonal and only to first order. Because off-diagonal entries also shift the orbital energies at second order, it can be lowered while the true orbital energies stay wrong, and increasing its weight indeed makes the gap worse (Sec.~\ref{sec:ablation}). Density matching uses the exact eigenpairs and focuses on the frontier, but is blind to the sectors of Proposition~\ref{prop:blind}.

\paragraph{Desiderata.}
A loss that reliably improves eigen-properties should therefore (i) respond to every sector, so that no error goes unpenalized; (ii) weight the sectors by their effect on the quantities of interest, so that model capacity is spent where it matters; and (iii) use the exact eigenpairs of $\Hh$ rather than a first-order proxy. No existing loss meets all three requirements. We now construct one that does.

\subsection{From the density to the cumulative LDOS}
\label{sec:loss}

The density is blind to orbital energies because it sums the occupied orbitals and discards their energies, as \Eqref{eq:density} shows. Keeping the energy of each orbital gives the local density of states (LDOS), $\rho(\mathbf{r},E)=\sum_i|\psi_i(\mathbf{r})|^2\,\delta(E-\varepsilon_i)$, where $E$ is an energy coordinate and $\delta(\cdot)$ is the Dirac delta, placing a peak at each orbital energy $\varepsilon_i$. The LDOS describes how the electrons at point $\mathbf{r}$ are distributed in energy and is measured directly by scanning tunnelling microscopy \citep{tersoff1985stm}. We use its cumulative form, with each peak smoothed by a Gaussian of width $\sigma$, so that each step becomes an error-function step:
\begin{equation}
\Nre(\mathbf{r},E)=\sum_{i=1}^{n}|\psi_i(\mathbf{r})|^2\,\Phi\!\Big(\frac{E-\varepsilon_i}{\sigma}\Big),
\label{eq:cumldos}
\end{equation}
where $\Phi$ is the standard normal cumulative distribution function and $\sigma$ is a smoothing width. $\Nre(\mathbf{r},E)$ is the number of electrons per spin at $\mathbf{r}$ with energy below $E$. As a function of $E$, it is a smoothed staircase with one step per orbital: the step sits at the orbital energy $\varepsilon_i$, and its height is the orbital density $|\psi_i(\mathbf{r})|^2$ at that point (Figure~\ref{fig:appspectral}b in the appendix). An error in an orbital energy moves a step (\textcolor{YellowOrange}{\textbf{A}}). An error in an orbital shape changes step heights, whether it mixes occupied with virtual orbitals (\textcolor{RoyalBlue}{\textbf{B}}) or orbitals within the occupied or the virtual space (\textcolor{violet}{\textbf{C}}); Figure~\ref{fig:teaser}f shows an example of \textcolor{YellowOrange}{\textbf{A}} and one of \textcolor{violet}{\textbf{C}}.

The staircase is what a scanning tunnelling microscope traces as its bias voltage is swept. With the tip held above a point $\mathbf r$, the tunnelling current at bias $V$ counts the states between the Fermi level $E_F$ and $E_F+eV$, so it is proportional to $\Nre(\mathbf r,E_F+eV)-\Nre(\mathbf r,E_F)$ \citep{tersoff1985stm,selloni1985stm}.\footnote{A short video on scanning tunnelling microscopy (La Physique Autrement): \url{https://youtu.be/HE2yE8SvHmA?t=73}.} Each time the bias crosses an orbital energy, the current changes by one step, whose height is that orbital's density at the tip. The density is one value of $\Nre$, its level at the Fermi energy. The cumulative LDOS loss also compares how $\Nre$ changes across the energy window, which is what the bias sweep traces, at many points $\mathbf r$, much as one would compare two samples by sweeping the bias at many tip positions.

Where the density matrix, and with it the density, keeps only the occupied subspace, $\Nre$ keeps the energy and the spatial density of every orbital, occupied and virtual: the positions and heights of its steps. These are enough to rebuild the Hamiltonian, $\mathbf H=\Sm\mathbf C\diag(\epsv)\mathbf C^\top\Sm$, so the map $\mathbf H\mapsto\Nre[\mathbf H]$ discards nothing (Theorem~\ref{thm:props}(iii)). At an energy inside the HOMO--LUMO gap, every occupied step is complete and no virtual step has begun, so that slice of $\Nre$ is the density, up to exponentially small tails (Theorem~\ref{thm:props}(i)). $\Nre$ is a single function of the predicted eigenpairs, not a sum of terms for the three sectors: the sectors are only the lens through which Section~\ref{sec:props} shows that it responds to every part of the error.

\begin{definition}[Cumulative LDOS loss]
Let $\mathcal{R}$ be a set of sample points, $\{E_k\}_{k=1}^{K}$ an energy grid on the window $[\mu-E_{\downarrow},\mu+E_{\uparrow}]$ around the reference Fermi level $\mu=\tfrac12(\varepsilon^\star_{n_o}+\varepsilon^\star_{n_o+1})$, and $\mathrm{div}$ the factor in \Eqref{eq:lrho}. The cumulative LDOS loss is
\begin{equation}
\Lcum(\Hh)=\Big\langle\Big(\frac{\Nre[\Hh](\mathbf{r},E_k)-\Nre[\Hs](\mathbf{r},E_k)}{\mathrm{div}(\mathbf{r})}\Big)^2\Big\rangle_{\mathbf{r}\in\mathcal{R},\,k},
\label{eq:lcum}
\end{equation}
where $\Nre[\Hh]$ is computed from the exact eigenpairs of $(\Hh,\Sm)$ and gradients are propagated through the eigensolver.
\end{definition}

FermiScope trains with $\Lelem+w\,\Lcum$, where $\Lelem$ is the element-wise loss of Section~\ref{sec:background} and the weight $w>0$ balances the scales of the two terms, selected on the validation split (Table~\ref{tab:hparams}); on QHNet, the results change little over a tenfold range of $w$ (Sec.~\ref{sec:ablation}). Two variants serve as ablations: replacing the step $\Phi$ by a Gaussian peak $e^{-x^2/2}$ gives a \emph{peak-shaped} LDOS loss $\Lldos$, and a single grid point at $E_1=\mu$ recovers density matching (Theorem~\ref{thm:props}(i)).

\subsection{Theoretical properties}
\label{sec:props}

\begin{theorem}[Properties of the cumulative LDOS]
\label{thm:props}
Let the reference orbital energies be non-degenerate, as in Proposition~\ref{prop:sectors}, let $g=\varepsilon^\star_{n_o+1}-\varepsilon^\star_{n_o}$ be the reference HOMO--LUMO gap, and let $I$ be any open energy interval, for example the loss window. Then:
\begin{enumerate}[label=(\roman*)]
\item \emph{Density matching is a special case.} If no orbital energy lies within $g/2$ of $\mu$, then $\Nre(\mathbf{r},\mu)=\rho(\mathbf{r})/2$ up to $\tfrac12\erfc\big(g/(2\sqrt2\sigma)\big)$ per orbital, so $\Lrho$ is the $E=\mu$ slice of $\Lcum$.
\item \emph{No blind sector.} The first-order response of $\Nre$ on $\mathbb{R}^3\times I$ to $\Dl$ is injective: every energy error $d_i$ (sector~\textcolor{YellowOrange}{A}) and every off-diagonal entry $\Delta_{ij}$ (sectors~\textcolor{RoyalBlue}{B} and \textcolor{violet}{C}) changes $\Nre$.
\item \emph{Identifiability.} $\Nre[\Hh]=\Nre[\Hs]$ on $\mathbb{R}^3\times I$ if and only if $\Hh=\Hs$.
\item \emph{Non-saturating gradient.} If one orbital energy is displaced by $\delta$ with its shape fixed, the gradient of the cumulative mismatch with respect to $\delta$ is proportional to $\erf(\delta/2\sigma)$, the Gauss error function\footnote{$\erf(x)=\frac{2}{\sqrt\pi}\int_0^x e^{-t^2}\,dt$; it increases monotonically from $-1$ to $+1$.}, which saturates at $\pm1$ as $|\delta|$ grows. In contrast, the gradient of the peak-shaped mismatch is proportional to $(\delta/\sigma)\,e^{-\delta^2/4\sigma^2}$, which vanishes for large displacements. Concretely, at $|\delta|=4\sigma$ the cumulative gradient has reached $\erf(2)>0.99$ of its asymptotic value while the peak gradient has fallen to $4e^{-4}<0.08$, a factor of 13 apart, and the ratio grows without bound as $|\delta|$ increases.
\end{enumerate}
\end{theorem}

Appendix~\ref{app:proofs} proves each part (Propositions~\ref{prop:identity} and~\ref{prop:saturation}, Theorem~\ref{thm:complete}) for any linearly independent basis of Gaussian-type orbitals, and Appendix~\ref{app:verify} checks every part numerically on QH9 molecules.

\begin{figure}[t]
\centering
\includegraphics[width=\textwidth]{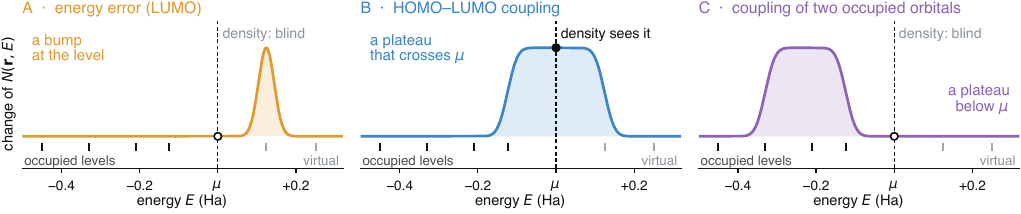}
\vspace{-1.4em}
\caption{How each kind of error shows up in the cumulative LDOS: the first-order change of $\Nre(\mathbf r,E)$ at one point $\mathbf r$ as a function of energy (schematic levels with the median QH9 gap, $\sigma=0.02$\,Ha; sign and scale arbitrary). An orbital-energy error (\textcolor{YellowOrange}{\textbf{A}}, left) adds a bump at its level. An occupied--virtual coupling (\textcolor{RoyalBlue}{\textbf{B}}, middle) adds a plateau between the two levels, which crosses the Fermi level $\mu$. A coupling of two occupied orbitals (\textcolor{violet}{\textbf{C}}, right) adds a plateau that ends below $\mu$. Density matching reads only the slice $E=\mu$ (dashed line), where only the \textcolor{RoyalBlue}{\textbf{B}} plateau is nonzero; the cumulative LDOS loss reads the whole energy window and sees all three (Theorem~\ref{thm:props}(ii)).}
\label{fig:fingerprints}
\vspace{-0.8em}
\end{figure}

Figure~\ref{fig:fingerprints} shows why (ii) holds and why the density fails it. To first order, an energy error adds a bump to $\Nre$ at its level, and a coupling $\Delta_{ij}$ adds a plateau between $\varepsilon^\star_i$ and $\varepsilon^\star_j$, shaped in space like $\psi_i\psi_j$. Steps and bumps at different energies are linearly independent functions of $E$, and different orbitals are linearly independent functions of $\mathbf r$, so no combination of errors can cancel. The density reads only the slice $E=\mu$. Bumps never reach it, and a plateau crosses it only if one of its two orbitals is occupied and the other virtual: this is Proposition~\ref{prop:blind} seen through $\Nre$.

\paragraph{Implications.}
By (i), which holds at every energy inside the gap and hence at the two grid energies next to $\mu$, the cumulative loss keeps everything that density matching sees, including its automatic focus on the frontier orbitals (Proposition~\ref{prop:sectors}c); the other energies add the orbital-energy information that the density lacks. By (ii) and (iii), all three sectors are visible, and on the loss window $\Nre$ determines the Hamiltonian, so it cannot be matched by trading one kind of error for another, as the diagonal-only WALoss can. Pairs of orbitals that both lie far outside the window, such as two core orbitals or two high virtual orbitals, enter only through the Gaussian tails of their energy profiles; they barely affect ground-state properties and remain supervised by the element-wise term. The loss therefore concentrates on the valence and low virtual orbitals, the part of the spectrum that governs chemistry. Together, these properties meet the three desiderata of Sec.~\ref{sec:analysis}, and (iv) makes training robust: unlike a peak-shaped loss, the cumulative loss keeps pulling an orbital energy toward its target once the error exceeds a few $\sigma$, because it compares the local spectra through their cumulative distributions, as a transport distance does, rather than through the overlap of peaks (Proposition~\ref{prop:saturation} in Appendix~\ref{app:proofs}).

Finally, because most degrees of freedom of $\Dl$ lie in sectors that barely affect the target properties, a model can lower $\Lcum$ by moving error out of the sensitive sectors rather than by enlarging it, so the element-wise error need rise only slightly (Proposition~\ref{prop:realloc} in Appendix~\ref{app:proofs}).

\section{Experiments}
\label{sec:exp}

\subsection{Setup}
\label{sec:setup}

\paragraph{Datasets.}
QH9 \citep{yu2023qh9} provides B3LYP/def2-SVP Hamiltonians for QM9 molecules \citep{ramakrishnan2014qm9} with up to nine heavy atoms. We use its four official splits: \emph{stable-id} is a random split, \emph{stable-ood} tests on larger molecules, \emph{dynamic-geo} on unseen geometries of training molecules and \emph{dynamic-mol} on unseen molecules. Appendix~\ref{app:data} gives sizes and statistics.

\paragraph{Backbones.}
We fine-tune three pretrained Hamiltonian networks that span the development of the field: QHNet \citep{yu2023qhnet}, the original QH9 baseline, and WANet \citep{li2025wanet} and SPHNet \citep{luo2025sphnet}, which make equivariant message passing cheaper through SO(2) convolutions and adaptive sparsity.

\paragraph{Baselines and protocol.}
We compare with WALoss \citep{li2025wanet}, the state-of-the-art physics-informed loss for Hamiltonian prediction, in its full form. For each backbone, every method fine-tunes the same pretrained checkpoint for 30k steps with the same optimizer, learning-rate schedule, batch size and data. Only the auxiliary loss and its weight differ, and all methods are evaluated with the same code. Appendix~\ref{app:impl} lists all hyper-parameters.

\paragraph{Training objective and cost.}\label{sec:impl}
We apply $\Lcum$ when fine-tuning a pretrained model. At each step it is evaluated on eight molecules of the batch, at 256 sample points per molecule drawn once from the PySCF quadrature grid \citep{sun2018pyscf}, whose atomic-orbital values are cached, and on $K=64$ energies on $[\mu-1.5,\mu+0.3]$\,Ha with $\sigma=0.02$\,Ha, covering the valence occupied orbitals, the LUMO and the low virtual orbitals. The eigenpairs of $(\Hh,\Sm)$ are computed in double precision, with gradients through the eigensolver (Appendix~\ref{app:lossimpl}). On QHNet the loss adds about 5\% to the cost of pretraining, and it costs nothing at inference (Appendix~\ref{app:cost}).

\paragraph{Metrics.}
Following QH9, we report the MAE of $\Hh$ over all entries ($H$), over same-atom blocks (diag) and over the remaining blocks (non-diag). We also report the MAE of the HOMO and LUMO energies, of the gap and of all occupied orbital energies ($\epsv_{\occ}$), and the cosine similarity $\psi$ of the occupied orbital coefficients. Energies are in \uHa\ and $\psi$ in \% (Appendix~\ref{app:metrics}).

\subsection{Main results}
\label{sec:main}

\begin{table}[t]
\centering
\caption{Results on the QH9-stable-id test set (13{,}084 molecules) after 30k fine-tuning steps. Energies in \uHa, $\psi$ in \%. For each backbone, the gray rows are the two baselines, the pretrained model and fine-tuning with WALoss \citep{li2025wanet}. The violet row, \emph{density only}, is the ablation at weight $w=100$ that keeps only the $E=\mu$ slice of our loss, which is density matching (Proposition~\ref{prop:identity}). The blue row is our cumulative LDOS loss, which adds the energy resolution. All fine-tuning runs share every other setting, and the peak-shaped variant is in Appendix~\ref{app:gauss}. Bold marks the best downstream metric per backbone.}
\label{tab:main}
\label{tab:ablation}
\small
\setlength{\tabcolsep}{4pt}
\resizebox{\textwidth}{!}{%
\begin{tabular}{lllrrr?rrrrr}
\toprule
backbone & role & method & $H$ & diag & non-diag & HOMO $\downarrow$ & LUMO $\downarrow$ & gap $\downarrow$ & $\epsv_{\occ}$ $\downarrow$ & $\psi$ $\uparrow$ \\
\midrule
\rowcolor{rowbase}\cellcolor{white} & & pretrained & 77.56 & 129.25 & 73.68 & 1041 & 9445 & 9632 & 1206 & 96.28 \\
\rowcolor{rowbase}\cellcolor{white} & \multirow{-2}{*}{baseline} & + WALoss & 78.04 & 119.03 & 74.95 & 847 & 5458 & 5835 & 570 & 96.79 \\
\cmidrule(lr){2-11}
\rowcolor{rowabl}\cellcolor{white} & ablation & + density only ($w{=}100$) & 76.82 & 118.70 & 73.65 & 719 & 1575 & 2081 & 356 & 97.87 \\
\rowcolor{rowours}\cellcolor{white}\multirow{-4}{*}{QHNet} & ours & + cumulative LDOS (ours) & 74.81 & 109.54 & 72.19 & \textbf{678} & \textbf{1482} & \textbf{2027} & \textbf{341} & \textbf{97.91} \\
\midrule
\rowcolor{rowbase}\cellcolor{white} & & pretrained & 36.60 & 44.79 & 35.38 & 345 & 1309 & 1554 & 380 & 99.06 \\
\rowcolor{rowbase}\cellcolor{white} & \multirow{-2}{*}{baseline} & + WALoss & 37.28 & 46.03 & 35.98 & 341 & 957 & 1259 & 288 & 99.17 \\
\cmidrule(lr){2-11}
\rowcolor{rowabl}\cellcolor{white} & ablation & + density only ($w{=}100$) & 37.63 & 47.14 & 36.23 & 363 & 684 & 970 & 143 & 99.08 \\
\rowcolor{rowours}\cellcolor{white}\multirow{-4}{*}{SPHNet} & ours & + cumulative LDOS (ours) & 36.64 & 44.83 & 35.42 & \textbf{292} & \textbf{572} & \textbf{789} & \textbf{137} & \textbf{99.25} \\
\midrule
\rowcolor{rowbase}\cellcolor{white} & & pretrained & 40.07 & 51.28 & 38.42 & 335 & 1454 & 1684 & 288 & 98.98 \\
\rowcolor{rowbase}\cellcolor{white} & \multirow{-2}{*}{baseline} & + WALoss & 40.99 & 53.26 & 39.19 & 328 & 1025 & 1306 & 157 & 99.13 \\
\cmidrule(lr){2-11}
\rowcolor{rowabl}\cellcolor{white} & ablation & + density only ($w{=}100$) & 40.21 & 51.63 & 38.53 & 324 & 823 & 1101 & 140 & \textbf{99.22} \\
\rowcolor{rowours}\cellcolor{white}\multirow{-4}{*}{WANet} & ours & + cumulative LDOS (ours) & 40.19 & 51.59 & 38.52 & \textbf{287} & \textbf{573} & \textbf{791} & \textbf{137} & \textbf{99.22} \\
\bottomrule
\end{tabular}}
\vspace{-1em}
\end{table}

Table~\ref{tab:main} reports results on the QH9-stable-id test set. On QHNet, SPHNet and WANet, the cumulative LDOS loss gives the best orbital energies and orbital coefficients. It reduces the HOMO--LUMO gap error by 79\% on QHNet, 49\% on SPHNet and 53\% on WANet, and the LUMO error by 56--84\%; its gap error is 1.6--2.9 times lower than that of WALoss. These gains come without a cost in element-wise accuracy: the Hamiltonian MAE stays within 0.3\% of the pretrained value on SPHNet and WANet, and on QHNet it falls by 3.5\%, making our loss the best in every downstream column. WALoss improves the gap by 19--39\% but raises the Hamiltonian MAE on all three backbones (by 0.6--2.3\%), consistent with the trade-off that \citet{li2025wanet} reported on QH9.

This pattern follows from the analysis of Sec.~\ref{sec:analysis}. WALoss weights all entries of $\Dl$ equally, so much of its effect goes to the large sectors that barely influence the spectrum, and its spectral gains come at the expense of element-wise accuracy. The cumulative LDOS loss concentrates on the sectors that control the frontier spectrum, and the model can satisfy it by moving error out of these sectors at little cost in total error (Proposition~\ref{prop:realloc} in Appendix~\ref{app:proofs}).

\subsection{Generalization across splits}
\label{sec:generalization}

\begin{figure}[t]
\centering
\includegraphics[width=\textwidth]{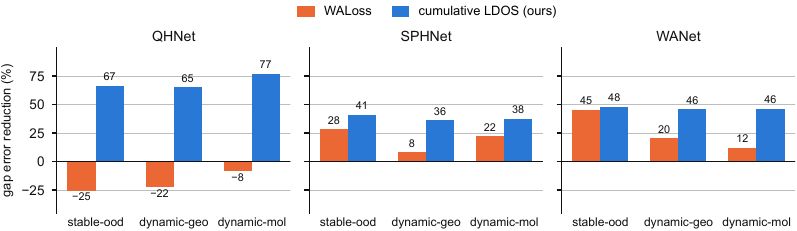}
\vspace{-1.5em}
\caption{Reduction of the HOMO--LUMO gap error relative to each pretrained model on the three shifted QH9 splits (higher is better; negative values mean that the error grew). The cumulative LDOS loss (ours, trained as $\Lelem+w\,\Lcum$) lowers the gap error on every backbone and split, and it also improves every other downstream metric; Table~\ref{tab:splits} in Appendix~\ref{app:splits} lists all metrics.}
\label{fig:splits}
\vspace{-1em}
\end{figure}

Figure~\ref{fig:splits} applies the same protocol to the three shifted splits; Table~\ref{tab:splits} in Appendix~\ref{app:splits} lists all metrics. The gains persist under distribution shift. Across all four splits, the cumulative LDOS loss lowers the gap error of every model, by 36--77\% on the shifted splits, and never raises the element-wise error by more than 0.3\%, whereas WALoss moves both errors in either direction and increases the gap error on all three shifted splits of QHNet. On dynamic-mol, where the pretrained QHNet is least accurate, our loss lowers its gap error by 77\% and raises its orbital similarity from 79.7\% to 90.6\%. Predicted Hamiltonians are also used to initialize DFT; starting SCF from them saves about a third of the iterations relative to the standard \texttt{minao} guess (Appendix~\ref{app:scf}).

\subsection{Ablation studies}
\label{sec:ablation}

\paragraph{Energy resolution.}
The ablation rows of Table~\ref{tab:main} isolate the first design choice of the loss, the energy resolution; the second, step versus peak, is in Appendix~\ref{app:gauss}. Density matching alone already lowers the gap error by 35--78\%, which confirms the value of its automatic focus on the frontier (Proposition~\ref{prop:sectors}c). Adding energy resolution helps most where the density is blind to the remaining error: on WANet the gap error falls from 1101 to 791\,\uHa, and on SPHNet from 970 to 789\,\uHa. On QHNet, where density matching already captures most of the gap improvement, energy resolution removes its element-wise cost ($H$ from 76.82 to 74.81\,\uHa) and further lowers the LUMO and occupied-orbital errors.

\paragraph{Weight sensitivity.}
On QHNet, the gap error of the cumulative loss stays between 1811 and 2231\,\uHa\ as $w$ varies tenfold, whereas the diagonal-only form of WALoss, which sees orbital energies only to first order, degrades as its weight grows (Appendix~\ref{app:dose}).

\section{Related Work}
\label{sec:related}

\paragraph{Hamiltonian prediction.}
SchNOrb \citep{schutt2019schnorb} and PhiSNet \citep{unke2021phisnet} predict molecular Hamiltonians and wavefunctions, and DeepH \citep{li2022deeph} targets materials. QHNet \citep{yu2023qhnet} introduced an efficient equivariant architecture, and QH9 \citep{yu2023qh9} established the benchmark used here. WANet \citep{li2025wanet} and SPHNet \citep{luo2025sphnet} improve accuracy and scalability through SO(2) convolutions and adaptive sparsity. All of them are trained with element-wise losses; our loss applies to each of them without modification.

\paragraph{Physics-informed and spectral objectives.}
\citet{gu2022eigen} regress eigenvalues directly, which requires differentiating through an eigensolver at every step, supervises the energies but not the orbitals, and was found hard to optimize on large systems \citep{li2025wanet}. WALoss \citep{li2025wanet} avoids the eigensolver by evaluating $\Hh$ in the reference orbital basis. Lemma~\ref{lem:whitening} shows that its full form is the element-wise error measured in an orthonormal basis, and its perturbation theorem bounds eigenvalue errors by that error. Our analysis refines this bound into a map of which entries matter for which property, and explains the trade-off between matrix error and properties that \citet{li2025wanet} observed on QH9. Self-consistency training \citep{zhang2024selfconsistency} rebuilds a Hamiltonian from the predicted density through the exchange--correlation functional and matches it to the prediction on unlabeled molecules; it enforces the SCF fixed point rather than the spectrum and is complementary to our loss.

\paragraph{Electron density and LDOS in machine learning.}
DeepDFT \citep{jorgensen2022deepdft}, ChargE3Net \citep{koker2024charge3net} and LAGNet \citep{ushenin2025lagnet} predict the electron density as an output; LAGNet introduced the sampling grid and core suppression that we use. \citet{ellis2021ldos} predict the LDOS itself as the target of a finite-temperature DFT surrogate for materials. We use the LDOS as a training signal for a predicted Hamiltonian and show why its cumulative form is the one to optimize.

\section{Conclusion}
\label{sec:conclusion}
The element-wise error of a predicted Kohn--Sham Hamiltonian is a poor guide to its eigen-properties, because each property depends on a specific part of the error. Representing the error in the reference orbital basis makes this dependence explicit and shows what existing physics-informed losses miss. The cumulative LDOS represents the Hamiltonian in the language of the density without losing the orbital energies: it sees every part of the error, weights it by its effect on the spectrum, and determines the Hamiltonian uniquely. As the drop-in training signal of FermiScope, it reduces the HOMO--LUMO gap error by 49--79\% on three architectures, maintains these gains under distribution shift, and leaves the element-wise accuracy intact. Appendix~\ref{app:limitations} discusses limitations and future work.

\subsection*{AI use statement}

In this work, we used generative AI tools to implement methods: AI coding
assistants helped write parts of the software implementation, including
experiment orchestration scripts and figure-generation code.  We have not used
generative AI tools to generate synthetic data sets, help develop theoretical
models or conceptual frameworks, formulate mathematical claims, provide critical
ingredients for proving mathematical claims, propose or refine hypotheses,
design or provide feedback on research methodology or experiments, assist with
translation, clean and reformat the dataset, support qualitative and thematic
data analysis, or interpret results.  All mathematical proofs were written by
the authors; generative AI tools were subsequently used to polish the
exposition, and numerical scripts were used to verify the correctness of every
closed-form expression (see Appendix~\ref{app:verify}).  Additionally, we used
generative AI tools to identify and summarize relevant literature during the
related-work survey and to edit the paper to improve readability after the
authors had written the full draft.  We have reviewed all AI-assisted work.
AI-generated code was tested for correctness and verified against quantum-chemical reference calculations by the authors.  Literature identified
by AI was verified through manual inspection of the original sources.  All
experiments were designed and executed by the authors, and all reported numbers
were checked against the DFT reference data.  We take responsibility for the
final content of this work, including text, claims, and artifacts produced with
the aid of generative AI.

\subsection*{Reproducibility statement}
All code for computing the proposed losses, the cached quadrature grids, and the evaluation scripts are provided in the supplementary material.
Appendix~\ref{app:impl} lists the fine-tuning settings and the evaluation protocol, and the README of the supplementary code gives the training commands with these settings.
Complete proofs of the theoretical results appear in Appendix~\ref{app:proofs}, and Appendix~\ref{app:verify} reports numerical checks of every analytical statement against reference data.
The three backbone architectures we build on, QHNet, SPHNet and WANet, are publicly available together with the QH9 dataset and its published splits.

\subsection*{Ethics statement}
This work develops auxiliary training objectives for neural-network models of molecular Kohn--Sham Hamiltonians.
The research uses only publicly available quantum-chemical datasets (QH9) and involves no human subjects, private data, or dual-use applications.
We are not aware of any direct negative societal impact; by improving the physical accuracy of predicted Hamiltonians without requiring additional data or architectural changes, the method may help reduce the computational cost of molecular simulations.

\subsection*{Acknowledgments}
This work was partially supported by NSF 2303037, NSF 2312501, NSF 2531008, SRC JUMP 2.0 Center, CDSC Industry Partners, Amazon Research Awards, Snapchat, and Google Gifts.

\newpage 

\bibliography{refs}
\bibliographystyle{preprint}

\clearpage
\appendix
\raggedbottom
\let\appendixsection\section
\renewcommand{\section}{\clearpage\appendixsection}
\section{Kohn--Sham DFT in Brief}
\label{app:ks}
\begin{figure}[H]
\centering
\includegraphics[width=\textwidth]{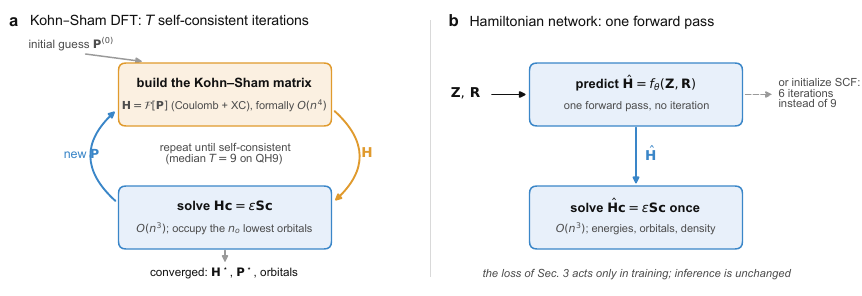}
\caption{\textbf{(a)} The self-consistent-field loop of Kohn--Sham DFT: the Kohn--Sham matrix is built from the current density matrix and diagonalized, and the occupied orbitals give the next density matrix, until the two agree. \textbf{(b)} A Hamiltonian network predicts the converged matrix in one forward pass; one diagonalization gives the orbital energies, orbitals and density, and the prediction can also initialize SCF. Iteration counts are medians over 50 QH9-stable-ood test molecules (PySCF, B3LYP/def2-SVP).}
\label{fig:appscf}
\end{figure}

This section collects the quantum-chemistry background assumed in the main text: a glossary of terms (Table~\ref{tab:glossary}), and the self-consistent-field procedure and its cost (Figure~\ref{fig:appscf}).

\paragraph{Glossary.}
Table~\ref{tab:glossary} defines the quantum-chemistry terms used in the paper.

\begin{table}[H]
\centering\small
\caption{Quantum-chemistry terms used in this paper.}
\label{tab:glossary}
\renewcommand{\arraystretch}{1.12}
\begin{tabular}{@{}p{0.25\textwidth}p{0.71\textwidth}@{}}
\toprule
term & meaning \\
\midrule
atomic number $Z$ & Number of protons of a nucleus; it identifies the element (H 1, C 6, N 7, O 8, F 9). \\
basis set, basis function, atomic orbital & A fixed set of functions $\phi_\mu$ centred on the atoms, in which the molecular orbitals are expanded; each function is called an atomic orbital. \\
def2-SVP & A standard Gaussian basis set of split-valence quality with polarization functions; it gives 5 functions per H atom and 14 per C, N, O or F atom. \\
molecular orbital $\psi_i$ & A one-electron wavefunction of the molecule, a linear combination of basis functions; $|\psi_i(\mathbf{r})|^2$ is the probability density of an electron in that orbital. \\
Kohn--Sham DFT & Density functional theory in which the interacting electrons are replaced by non-interacting electrons in an effective potential that depends on the electron density. \\
exchange--correlation functional, B3LYP & The approximate part of the effective potential that accounts for many-electron effects; B3LYP is a widely used hybrid functional, which mixes in exact exchange. \\
level of theory & The functional and basis set of a calculation, here B3LYP/def2-SVP. \\
Kohn--Sham matrix $\mathbf{H}$ & The matrix of the Kohn--Sham operator in the basis, also called the Hamiltonian or Fock matrix. \\
overlap matrix $\mathbf{S}$ & The inner products of the basis functions; it differs from the identity because the basis functions are not orthogonal. \\
generalized eigenvalue problem & $\mathbf{Hc}=\varepsilon\mathbf{Sc}$; its eigenvalues are the orbital energies and its eigenvectors the orbital coefficients. \\
self-consistent field (SCF) & The iteration that makes the Hamiltonian and the density consistent with each other (Figure~\ref{fig:appscf}). \\
initial guess (\texttt{minao}, core) & The starting point of SCF: \texttt{minao} superposes atomic densities in a minimal basis; the core guess ignores the interaction between electrons. \\
DIIS & The standard extrapolation scheme that accelerates SCF convergence \citep{pulay1980diis}. \\
closed shell & Every occupied orbital holds two electrons of opposite spin, so $n_o=N_e/2$. \\
occupied, virtual orbitals & The $n_o$ lowest orbitals hold the electrons; the others, called virtual or unoccupied, are empty. \\
HOMO, LUMO, gap & The highest occupied and lowest unoccupied molecular orbitals, and the difference of their energies, a common indicator of reactivity and optical excitation. \\
frontier orbitals & The orbitals closest in energy to the HOMO--LUMO gap. \\
core orbitals & The deep $1s$ orbitals of the heavy atoms, far below the valence orbitals. \\
Fermi level $\mu$ & Here, the energy midway between the HOMO and the LUMO; the orbitals below it are occupied. \\
electron density $\rho(\mathbf{r})$ & The expected number of electrons per unit volume at $\mathbf{r}$. \\
density matrix $\mathbf{P}$ & $2\mathbf{C}_{\occ}\mathbf{C}_{\occ}^\top$, the matrix that gives the density through $\rho(\mathbf{r})=\boldsymbol{\phi}(\mathbf{r})^\top\mathbf{P}\,\boldsymbol{\phi}(\mathbf{r})$. \\
local density of states (LDOS) & The density at $\mathbf{r}$ resolved by orbital energy; it is measured by scanning tunnelling microscopy. \\
units & Hartree (Ha), the atomic unit of energy: 1\,Ha $=27.211$\,eV, 1\,\uHa\ $=10^{-6}$\,Ha, 1\,kcal/mol $=1.594$\,mHa. \\
\bottomrule
\end{tabular}
\end{table}

\clearpage
\paragraph{The SCF procedure and its cost.}
The Kohn--Sham operator $\mathcal{F}[\Pm]$ depends on the occupied orbitals through their density matrix, and the density matrix is computed from the eigenvectors of its matrix. DFT resolves this circularity by iteration (Figure~\ref{fig:appscf}a). Starting from an initial guess of the density matrix, each iteration builds the Kohn--Sham matrix from the current density matrix, solves the generalized eigenvalue problem, \Eqref{eq:gep}, occupies the $n_o$ lowest orbitals and forms a new density matrix by \Eqref{eq:density}, until the input and output density matrices agree. Building the matrix requires the Coulomb and, for hybrid functionals such as B3LYP, the exact-exchange contributions, formally $O(n^4)$ in the number of basis functions, together with a numerical quadrature of the exchange--correlation potential; the diagonalization costs $O(n^3)$. A calculation with $T$ iterations costs $T$ times one iteration. On 50 QH9-stable-ood test molecules, PySCF with its default settings (B3LYP/def2-SVP, DIIS \citep{pulay1980diis}) converges in a median of $T=9$ iterations (range 8--10) from its default \texttt{minao} guess and in 17 from the core-Hamiltonian guess, taking a median of about 9\,s per molecule on a CPU node. A Hamiltonian network replaces the whole loop (Figure~\ref{fig:appscf}b): one forward pass predicts the $O(n^2)$ entries of $\Hh$, and one diagonalization gives every downstream quantity. On the same molecules, the QHNet forward pass on a GPU and the eigensolve take a median of 0.1\,s together. Used instead as an initial guess, the prediction saves about a third of the SCF iterations (Appendix~\ref{app:scf}). The loss proposed in this paper changes only training; inference is that of the backbone.

\input{app-theory}

\section{The QH9 Datasets and Splits}
\label{app:data}

\subsection{Provenance and quantum-chemical settings}
Every experiment in this paper uses QH9 \citep{yu2023qh9}, which supplies converged Kohn--Sham Hamiltonians for QM9 \citep{ramakrishnan2014qm9} molecules. The reference calculations are B3LYP with the def2-SVP basis, carried out in PySCF \citep{sun2018pyscf}; each record stores the converged Fock matrix $\Hs$ in the atomic-orbital basis together with the overlap matrix $\Sm$, the nuclear charges and the geometry. Because $\Hs$ is the converged operator, the label is a fixed point of the self-consistency map and not a single SCF iterate, so a model that reproduces it reproduces the entire converged electronic structure of that geometry.

Two families are released. \emph{QH9-stable} holds one equilibrium geometry for each of 130{,}831 molecules. \emph{QH9-dynamic} holds short \emph{ab initio} molecular-dynamics trajectories; we use the 100k release, which is 999 molecules sampled at 100 geometries each, 99{,}900 records in total. The dynamic family is the harder of the two: consecutive frames of a trajectory visit distorted, off-equilibrium geometries whose Hamiltonians are not close to any equilibrium structure in the stable set.

The basis is small enough to state completely. In def2-SVP a hydrogen carries five atomic orbitals ($2s$, $1p$) and each of C, N, O and F carries fourteen ($3s$, $2p$, $1d$), so the matrix dimension is determined by the composition alone: $n=5n_{\mathrm{H}}+14n_{\mathrm{heavy}}$. The two molecules used for the figures below check this exactly: molecule $\#12345$ is C$_7$H$_{10}$O with $n=7\cdot14+1\cdot14+10\cdot5=162$, and molecule $\#23470$ is C$_8$H$_5$N with $n=9\cdot14+5\cdot5=151$.

\subsection{Splits}
The four splits used in this paper are given in Table~\ref{tab:splits_def}. They probe different kinds of generalization, which matters for the interpretation of Table~\ref{tab:splits}: \emph{stable-id} is an interpolation test, \emph{stable-ood} extrapolates in molecule size, \emph{dynamic-geo} extrapolates to unseen geometries of molecules the model has seen, and \emph{dynamic-mol} extrapolates to unseen molecules altogether. The size split is a substantial shift. Measured on a uniform 4{,}000-molecule sample of QH9-stable, the training side has a median of 17 atoms and $n=166$, while the test side has a median of 23 atoms and $n=196$, and the median gap moves from 6.43 to 8.10\,eV: the test molecules are systematically larger, and have wider HOMO--LUMO gaps, than the training molecules.

\begin{table}[H]
\centering\small
\caption{The four QH9 splits. Sizes are the exact record counts read from the released split masks. The size split thresholds are on the total atom count, hydrogens included; the random split uses a fixed permutation with seed 43.}
\label{tab:splits_def}
\footnotesize
\setlength{\tabcolsep}{3pt}
\begin{tabular}{llrrrl}
\toprule
split & selection rule & train & val & test & tests \\
\midrule
stable-id   & random $80/10/10$            & 104{,}664 & 13{,}083 & 13{,}084 & interpolation \\
stable-ood  & $\le20$ / $21$--$22$ / $\ge23$ atoms & 104{,}001 & 17{,}495 & 9{,}335 & molecule size \\
dynamic-geo & geometries within a molecule & 79{,}920 & 9{,}990 & 9{,}990 & unseen geometry \\
dynamic-mol & disjoint molecules           & 79{,}900 & 9{,}900 & 10{,}100 & unseen molecule \\
\bottomrule
\end{tabular}
\end{table}

\clearpage
\subsection{Composition, spectra, and the size of the blind subspace}
\begin{figure}[H]
\centering
\includegraphics[width=0.70\textwidth]{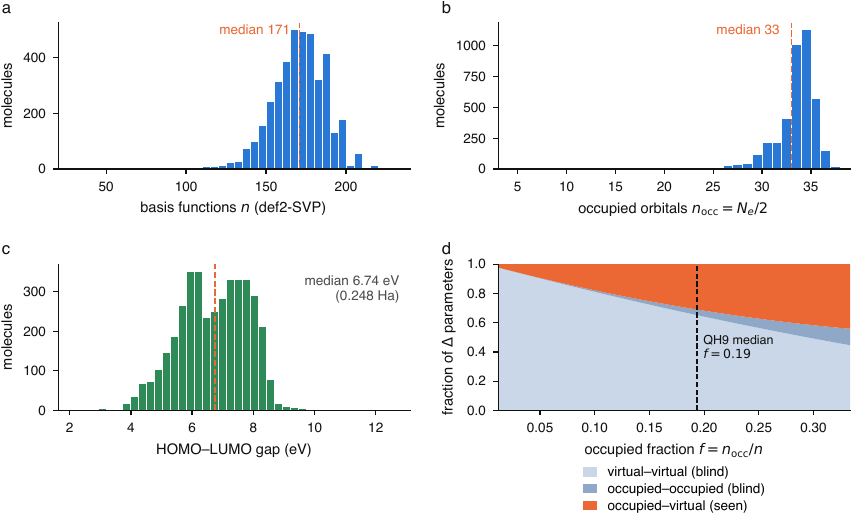}
\caption{QH9-stable statistics from a uniform 4{,}000-molecule sample. \textbf{(a)} Basis size. \textbf{(b)} Number of occupied orbitals. \textbf{(c)} Reference HOMO--LUMO gap. \textbf{(d)} Exact partition of the $n(n+1)/2$ parameters of a symmetric residual $\Dl$ into the three sectors of Proposition~\ref{prop:sectors}, at fixed $n=171$, as a function of the occupied fraction $f$. Only the occupied--virtual slice changes the density; at the QH9 median occupied fraction it accounts for $31\%$ of the parameters, and the remaining $69\%$ is invisible to $\Lrho$.}
\label{fig:appdata}
\end{figure}

Table~\ref{tab:datastats} and Figure~\ref{fig:appdata} summarize the sample. Two entries matter for the rest of the paper. First, the occupied fraction is small: the median molecule has $n_{\mathrm{occ}}=33$ occupied orbitals out of $n=171$ basis functions, so $f=n_{\mathrm{occ}}/n\approx0.19$. Second, the dimension count of Proposition~\ref{prop:blind} is therefore severe. A symmetric residual $\Dl$ has $n(n+1)/2$ free parameters, of which $n_{\occ}(n_{\occ}+1)/2+n_{\vir}(n_{\vir}+1)/2$ lie in the occupied--occupied and virtual--virtual blocks that the density cannot resolve. At the QH9 median that is a fraction of $0.690$, and the $5$--$95\%$ range across the sample is $[0.661,0.709]$: roughly seven of every ten directions in which a predicted Hamiltonian can be wrong are directions in which density matching is exactly blind, and this holds for every molecule in the sample.

The spectra are also wide. The median molecule has its lowest eigenvalue at $-19.1$\,Ha and its highest at $+3.7$\,Ha, while the quantities we care about live in a window of order 1\,Ha around the Fermi level. Nine deep core levels of molecule $\#23470$ sit below the loss window and thirty-two levels lie inside it; the cumulative form is defined so that the levels below the window contribute a constant step and drop out of any difference (Appendix~\ref{app:impl}).

\begin{table}[H]
\centering\small
\caption{QH9-stable composition and reference spectra, measured on a uniform 4{,}000-molecule sample. Ranges are $5$--$95\%$ intervals unless a min--max is stated. The last row is the exact fraction of the parameters of a symmetric residual that lie in the two sectors the density loss cannot resolve.}
\label{tab:datastats}
\begin{tabular}{lrrr}
\toprule
quantity & median & range & min--max \\
\midrule
atoms per molecule                    & 18     & --            & 5--29 \\
heavy atoms (C, N, O, F)              & 9      & --            & 1--9 \\
basis functions $n$                   & 171    & 141--196      & 34--226 \\
occupied orbitals $n_{\mathrm{occ}}$  & 33     & 29--35        & 5--37 \\
occupied fraction $f=n_{\mathrm{occ}}/n$ & 0.193 & 0.178--0.219 & 0.147--0.256 \\
HOMO--LUMO gap (Ha)                   & 0.248  & 0.173--0.305  & 0.079--0.463 \\
HOMO--LUMO gap (eV)                   & 6.74   & 4.72--8.31    & 2.15--12.60 \\
lowest eigenvalue $\varepsilon_1$ (Ha) & $-19.14$ & --         & $-24.76$--$-10.17$ \\
highest eigenvalue $\varepsilon_n$ (Ha) & $3.69$ & --          & $2.91$--$4.14$ \\
\midrule
density-blind fraction of $\Dl$       & 0.690  & 0.661--0.709  & -- \\
\bottomrule
\end{tabular}
\end{table}

\clearpage
\subsection{A gallery}
\begin{figure}[H]
\centering
\includegraphics[width=\textwidth]{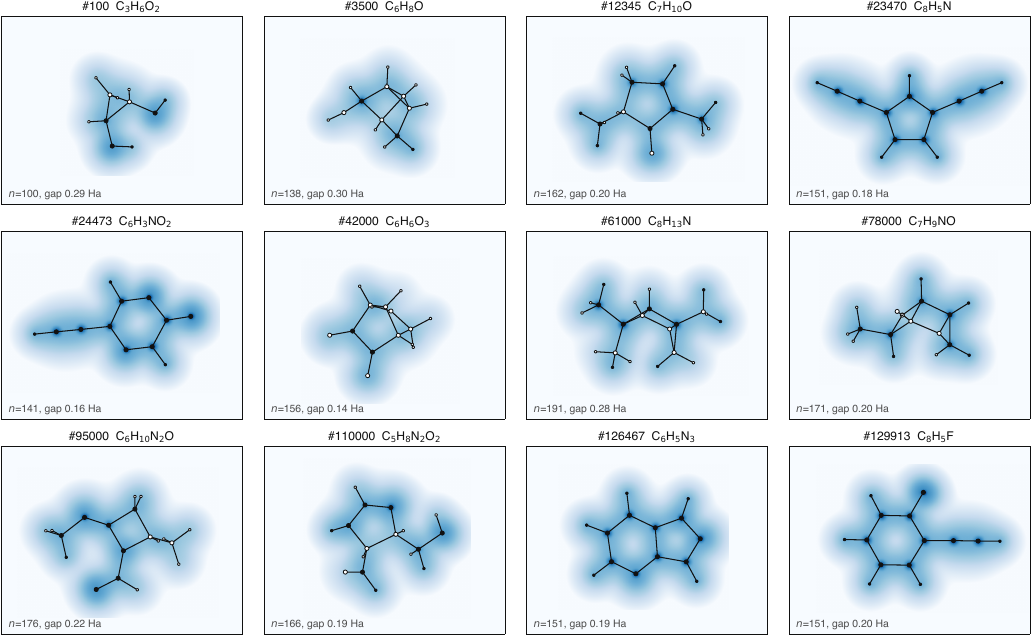}
\caption{Reference electron density for twelve QH9-stable molecules, on the best-fit plane through the nuclei, logarithmic colour scale, all panels on the same length scale and the same colour scale. Bonds are drawn from covalent-radius criteria; nuclei more than $0.7\,a_0$ off the plane are drawn open, so a panel with many open markers is a non-planar molecule seen in projection. Molecular formulae are in Hill notation.}
\label{fig:appgallery}
\end{figure}

Figure~\ref{fig:appgallery} shows the reference density of twelve molecules drawn at even spacing through the QH9-stable index range, on a common length scale. The sample is chemically diverse: saturated and conjugated skeletons, three- to six-membered rings, one to three heteroatoms, planar and strongly non-planar geometries, $n$ from 100 to 191 and reference gaps from $0.14$ to $0.30$\,Ha. A single auxiliary loss with a single weight has to be useful across all of it.

\subsection{The cached auxiliary quantities}
The auxiliary losses need real-space and spectral quantities that the released dataset does not contain, so we precompute them once per geometry and cache them. Each cache record holds the level-0 atom-centred radial--Lebedev quadrature grid of PySCF (14{,}112 points for molecule $\#23470$) with its weights, the reference density $\rho^\star$ on that grid, the geometry-only core-suppression factor $\mathrm{div}$, the fixed random subset $\mathcal{R}$ of 256 sample points, and the atomic-orbital values at those points. Caching the orbital values is what makes the loss cheap: evaluating $\Nre(\mathbf{r},E)$ during training costs one eigendecomposition and two small matrix products, with no basis-function evaluation in the training loop.

Two checks confirm the cache is consistent with the released labels. Rebuilding the def2-SVP molecule from the stored geometry and recomputing the overlap integrals reproduces the released $\Sm$ to $2.2\times10^{-7}$, and integrating the cached density against the cached weights gives $59.978$ electrons for a molecule with $N_e=60$, the expected accuracy of a level-0 grid.

\section{Visualizing the Reference Data}
\label{app:vis}
This section shows what the objects in Sections~\ref{sec:background} and \ref{sec:method} actually look like. Everything plotted here is reference DFT data: no learned model enters any panel. The purpose is to make the structural claims of the sector analysis concrete, and to make the scale of the quantities the losses compare explicit.

\subsection{The Hamiltonian and overlap matrices}
\begin{figure}[H]
\centering
\includegraphics[width=\textwidth]{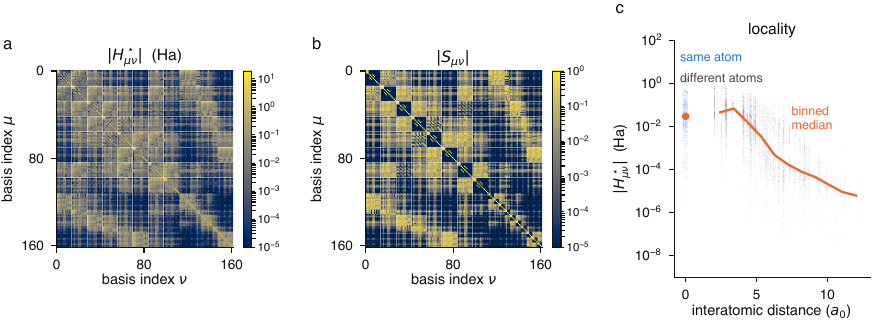}
\caption{Anatomy of a reference Kohn--Sham Hamiltonian, molecule $\#12345$ (C$_7$H$_{10}$O, $n=162$, $n_{\mathrm{occ}}=30$). \textbf{(a)} $|\Hs_{\mu\nu}|$ and \textbf{(b)} $|\Sm_{\mu\nu}|$ on a logarithmic colour scale; thin white rules mark heavy-atom block boundaries. Entries below $10^{-5}$ are shown at the floor of the scale. \textbf{(c)} Magnitude of the Hamiltonian elements against the distance between the atoms carrying the two basis functions, with same-atom pairs placed at zero and jittered. The orange curve is the binned median and the orange marker the same-atom median. The decay over six orders of magnitude is what local message-passing architectures exploit.}
\label{fig:apphamil}
\end{figure}

Figure~\ref{fig:apphamil} shows the two matrices for molecule $\#12345$. Three features shape the metrics reported throughout the paper. The dynamic range is large: $\max|\Hs_{\mu\nu}|=19.1$\,Ha, carried by the core diagonal, against a median magnitude three orders of magnitude smaller. The matrices are strongly local: the median $|\Hs_{\mu\nu}|$ is $3.0\times10^{-2}$\,Ha for basis functions on the same atom, $9.1\times10^{-3}$\,Ha for atoms $4$--$6\,a_0$ apart, and $1.9\times10^{-5}$\,Ha beyond $9\,a_0$. And the same-atom blocks dominate, which is why the diagonal-block MAE is roughly twice the off-diagonal-block MAE in every table in this paper; the split into ``diag'' and ``non-diag'' columns is a split by magnitude as much as by position.

This is also why element-wise accuracy is a weak proxy for spectral accuracy. An error budget dominated by a few large core entries can be reduced substantially while leaving the frontier of the spectrum untouched, and the element-wise-only control of Appendix~\ref{app:honly} does exactly that.

\clearpage
\subsection{Density, frontier orbitals, and core suppression}
\begin{figure}[H]
\centering
\includegraphics[width=\textwidth]{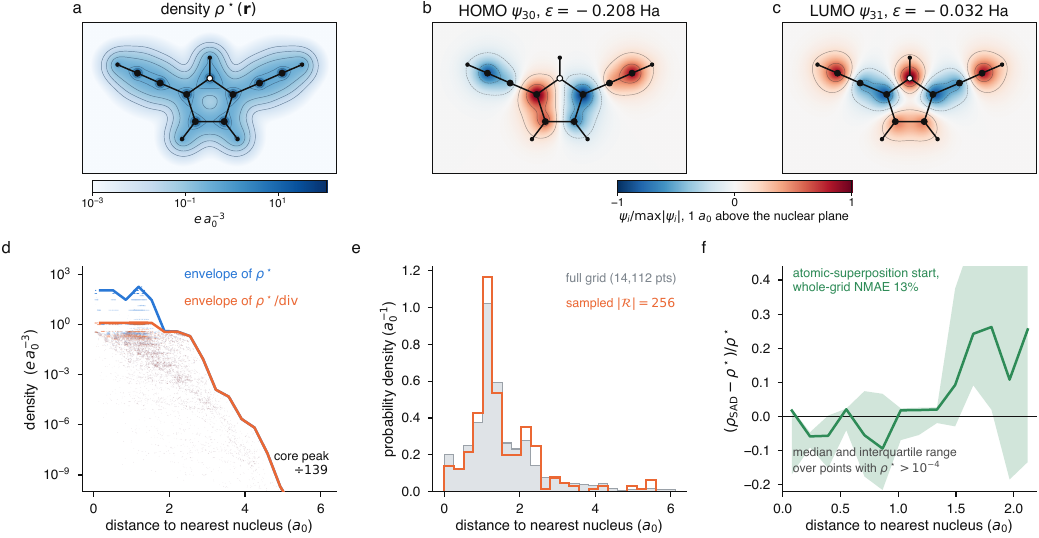}
\caption{Real-space reference quantities for molecule $\#23470$, 2,5-diethynyl-1H-pyrrole (C$_8$H$_5$N, $n=151$, $n_{\mathrm{occ}}=30$), which is exactly planar. \textbf{(a)} Electron density in the nuclear plane, logarithmic scale, with the molecular skeleton overlaid and the nitrogen drawn open. \textbf{(b,c)} HOMO and LUMO amplitudes sampled $1\,a_0$ above the nuclear plane and normalized by their own maxima; both vanish identically in the plane itself. \textbf{(d)} Upper envelope ($99.5$th percentile per radial bin) of $\rho^\star$ and of the core-suppressed target $\rho^\star/\mathrm{div}$ against distance to the nearest nucleus; the two coincide beyond $2\,a_0$, where $\mathrm{div}=1$. \textbf{(e)} The 256 sampled points reproduce the radial distribution of the full 14{,}112-point quadrature grid. \textbf{(f)} Relative error of the atomic-superposition density against distance to the nearest nucleus.}
\label{fig:apprealspace}
\end{figure}

Figure~\ref{fig:apprealspace} shows the real-space side for molecule $\#23470$, a planar molecule chosen so that the frontier orbitals can be drawn without projection artefacts. Both frontier orbitals are of $\pi$ symmetry: their amplitude in the nuclear plane is $1.8\times10^{-6}$, numerically zero, so they are sampled $1\,a_0$ above it. The HOMO is the bonding $\pi$ combination delocalized over the ring and both substituents; the LUMO adds a nodal plane across the ring. These are the objects that the occupied--virtual sector controls, and the reason the sector matters physically: an error in $\Dl_{\occ\vir}$ mixes the two pictures and moves both the gap and the frontier shapes.

Panel d explains the core-suppression factor. The reference density spans eighteen orders of magnitude, from $10^{-16}$ in the tails to $182\,e\,a_0^{-3}$ at the nuclei, and an unweighted squared residual would be a cusp-fitting objective. Dividing by the geometry-only factor $\mathrm{div}$ of \citet{ushenin2025lagnet}, which reaches $148$ at the nuclei and decays to exactly one beyond about $2\,a_0$, compresses the peak by a factor of $139$ and leaves the valence region untouched.
Panel f records how far an uninformed start is from the converged density: the atomic-superposition guess has a whole-grid NMAE of $13\%$, with the error concentrated in the bonding region around $1.5$--$2\,a_0$ where the atomic picture has nothing to say. Closing that gap is what buys the SCF savings in Table~\ref{tab:cost}.

\clearpage
\subsection{The spectral picture and the Fermi-level identity}
\begin{figure}[H]
\centering
\includegraphics[width=\textwidth]{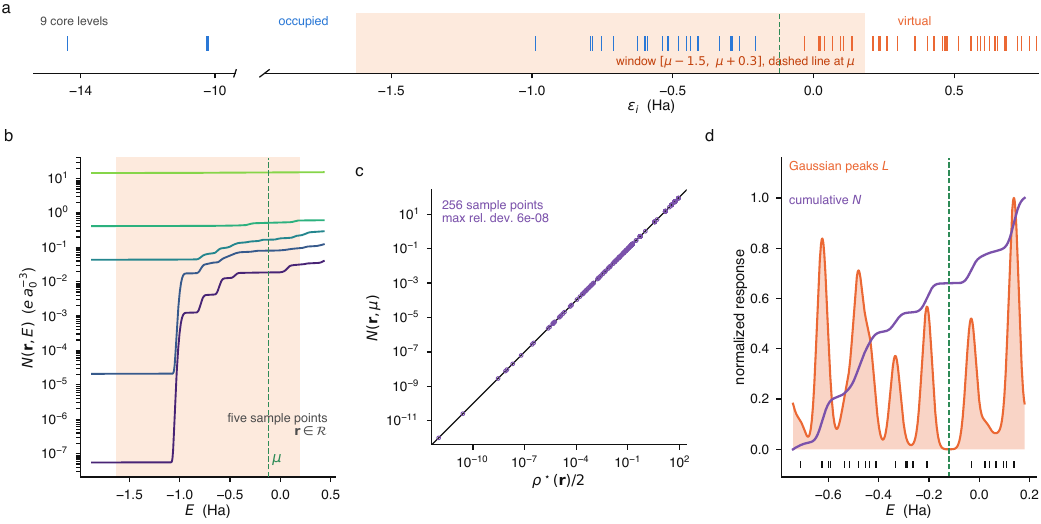}
\caption{Spectral reference quantities for molecule $\#23470$. \textbf{(a)} Generalized eigenvalues with the loss window shaded; the energy axis is broken so that the nine core levels are visible. \textbf{(b)} Cumulative LDOS $\Nre(\mathbf{r},E)$ of \Eqref{eq:cumldos} at five of the 256 sample points, spanning three orders of magnitude in local density. \textbf{(c)} Proposition~\ref{prop:identity} tested pointwise: $\Nre(\mathbf{r},\mu)$ against $\rho^\star(\mathbf{r})/2$ at all 256 sample points. \textbf{(d)} The two energy profiles at one sample point, with the constant contribution of the levels below the window removed and each curve normalized by its own maximum. Ticks mark the orbital energies.}
\label{fig:appspectral}
\end{figure}

Figure~\ref{fig:appspectral} shows the spectral side. Panel a places the energy window in context: nine core levels sit far below it, the valence occupied levels and the low virtuals lie inside, and the window width of $1.8$\,Ha is an order of magnitude larger than the gap it straddles. Panel b shows the cumulative LDOS at five sample points. Each curve is a staircase whose risers are the orbital weights $|\psi_i(\mathbf{r})|^2$ at that point, and the curves differ by orders of magnitude between points, which is why the loss is informative: a single molecule supplies 256 independent spectral fingerprints rather than one.

Panel c is a pointwise test of Proposition~\ref{prop:identity} on real data. Evaluating $\Nre(\mathbf{r},\mu)$ and $\rho^\star(\mathbf{r})/2$ independently at all 256 sample points agrees to a maximum relative deviation of $6\times10^{-8}$, which is the precision of the single-precision cached orbital values rather than of the identity. The cumulative loss therefore contains density matching as its $E=\mu$ slice, and the ablation between $\Lrho$ and $\Lcum$ in the main text compares one slice with the full profile.

Panel d shows why the extra slices help. At a fixed sample point the Gaussian-peak profile is a set of isolated bumps separated by intervals where it is numerically zero; a frontier orbital whose energy is wrong by more than a few $\sigma$ falls into one of those intervals and receives no gradient, which is the saturation of Proposition~\ref{prop:saturation} and the mechanism behind the deep-orbital drift in Table~\ref{tab:orbitals}. The staircase has no such intervals: it is monotone in $E$ and its plateaus still record how many states lie below, so the mismatch integrated over the window grows with the displacement instead of saturating.

\clearpage
\subsection{Where a real residual lives}
\label{app:realresidual}
\begin{figure}[H]
\centering
\includegraphics[width=\textwidth]{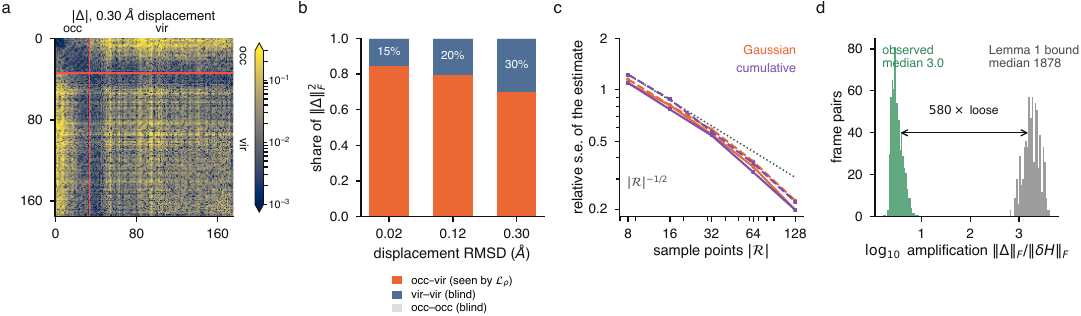}
\caption{Sector content of a physically realized Hamiltonian change, and two numerical properties of the estimator. \textbf{(a)} $|\Dl|$ for one pair of QH9-dynamic frames of the same molecule separated by $0.30$\,\AA\ RMSD, in the eigenbasis of the reference frame; red rules mark the occupied/virtual boundary at $n_{\mathrm{occ}}=34$, and values are clipped to the colour range at both ends. \textbf{(b)} Median share of $\|\Dl\|_F^2$ per sector against displacement, over 640 frame pairs from 214 trajectories. The virtual--virtual share, which contains the LUMO energy error and is invisible to $\Lrho$, doubles from $15\%$ to $30\%$ as the displacement grows. \textbf{(c)} Relative standard error of the per-molecule loss estimate when $|\mathcal{R}|$ of the 256 cached sample points are used, over 400 draws; solid lines are a random $75$\,\uHa\ element-wise residual, dashed lines a $10$\,mHa LUMO shift, and the grey guide is the $|\mathcal{R}|^{-1/2}$ rate. \textbf{(d)} Observed amplification $\|\Dl\|_F/\|\dH\|_F$ against the bound $\|\Sm^{-1}\|_2$ of Lemma~\ref{lem:whitening}, same frame pairs.}
\label{fig:appsector}
\end{figure}

The sector analysis of Section~\ref{sec:analysis} is a statement about an arbitrary residual. It is worth asking where a \emph{physically realized} residual actually sits, and the dynamic trajectories answer this without involving a model: taking two frames of the same molecule, $\dH=\Hs(\text{frame }k)-\Hs(\text{frame }0)$ is a genuine change of the converged Kohn--Sham operator caused by a geometric displacement, and $\Dl=\Cs_0{}^\top\dH\,\Cs_0$ is its whitened form in the eigenbasis of the reference frame. It is the kind of error a model makes when it fails to track a change of geometry, which is what the dynamic splits test. Figure~\ref{fig:appsector} reports the decomposition over 640 frame pairs drawn from 214 trajectories.

Three findings matter for the argument of this paper. First, the occupied--virtual sector dominates the magnitude: it carries a median of $70$--$84\%$ of $\|\Dl\|_F^2$ while occupying only $31\%$ of the parameters. Density matching is therefore well aimed, which is why $\Lrho$ alone already achieves a large share of the gap improvement in Table~\ref{tab:ablation}.

Second, the blind part is not negligible and it grows with the shift. At a displacement of $0.02$\,\AA\ the two density-blind sectors hold $15.6\%$ of $\|\Dl\|_F^2$; at $0.30$\,\AA\ they hold $30.3\%$, and essentially all of the growth is in the virtual--virtual block. That block is exactly where the frontier eigenvalue errors live: by Proposition~\ref{prop:sectors}(a) the LUMO energy error is $\Delta_{\mathrm{LUMO},\mathrm{LUMO}}$, a virtual--virtual entry, which Proposition~\ref{prop:blind} shows the density cannot resolve at any weight. The virtual--virtual share therefore grows exactly where a density-only signal becomes less informative, under geometric shift, which is the setting of the dynamic splits in Table~\ref{tab:splits}.

Third, the whitening bound of Lemma~\ref{lem:whitening} is loose by a wide margin on real data. The observed amplification $\|\Dl\|_F/\|\dH\|_F$ has a median of $3.0$ with a $5$--$95\%$ range of $[2.3,5.2]$, while the bound $\|\Sm^{-1}\|_2$ has a median of $1878$: the certificate overstates the true amplification by a median factor of $580$. Such a loose bound cannot certify that a small element-wise error implies a small spectral error; the looseness reflects the conditioning of def2-SVP overlap matrices, whose condition number has a median of $1.0\times10^{4}$ across QH9-stable. This is the quantitative version of the paper's starting observation: element-wise accuracy and spectral accuracy are only weakly coupled, so the spectrum has to be supervised directly.

\section{Metric Definitions}
\label{app:metrics}
All reported errors are test-set means over molecules, and all energies are converted to \uHa. Let $\Hh$ and $\Hs$ be the predicted and reference Hamiltonians of a molecule with $n$ basis functions and $n_{\mathrm{occ}}$ occupied orbitals, and let $\hat\epsv,\Ch$ and $\epsv^\star,\Cs$ solve the generalized eigenproblems $\Hh\mathbf{c}=\varepsilon\Sm\mathbf{c}$ and $\Hs\mathbf{c}=\varepsilon\Sm\mathbf{c}$ with the reference overlap $\Sm$ in both cases.

\paragraph{Element-wise errors.} $H$ is the mean absolute error over all matrix entries, $\frac{1}{n^2}\sum_{\mu\nu}|\Hh_{\mu\nu}-\Hs_{\mu\nu}|$. The \emph{diag} column restricts the average to entries whose two basis functions sit on the same atom, and \emph{non-diag} to the remaining entries; the two subsets partition the matrix, and by the locality shown in Figure~\ref{fig:apphamil}c the diagonal subset carries the larger entries. All three are invariant to any permutation of the basis, so they are comparable across code bases that order the atomic orbitals differently.

\paragraph{Orbital energies.} HOMO, LUMO and gap are the absolute errors of $\hat\varepsilon_{n_{\mathrm{occ}}}$, $\hat\varepsilon_{n_{\mathrm{occ}}+1}$ and their difference against the reference values. $\epsv_{\occ}$ is the mean absolute error over all occupied orbitals, $\frac{1}{n_{\mathrm{occ}}}\sum_{i\le n_{\mathrm{occ}}}|\hat\varepsilon_i-\varepsilon^\star_i|$, which is dominated by the deep levels and is therefore the column in which the deep-orbital drift of the peak-shaped loss (Table~\ref{tab:orbitals}) is visible. The per-orbital table (Table~\ref{tab:orbitals}) reports the same quantity resolved by orbital index relative to the frontier.

\paragraph{Orbital shapes.} $\psi$ is the mean cosine similarity between predicted and reference occupied orbitals, $\frac{1}{n_{\mathrm{occ}}}\sum_{i\le n_{\mathrm{occ}}}|\hat{\mathbf{c}}_i^\top\Sm\mathbf{c}^\star_i|$, reported as a percentage; the overlap metric is the natural inner product for the generalized problem, and the absolute value removes the sign ambiguity of an eigenvector. Values are high in absolute terms for every model, so the informative comparison is between methods within a backbone rather than against zero.

\paragraph{Downstream cost.} The SCF columns of Table~\ref{tab:cost} are ratios against a \texttt{minao} initialization on the same molecules and the same convergence threshold: iteration count, SCF wall time, and total wall time including the forward pass of the model. Ratios below one are savings.

\paragraph{Comparisons across backbones.} Each backbone is evaluated with its own code base. All comparisons in this paper are made within a backbone, between methods that share a checkpoint, a data split, an evaluation code and a step budget.

\section{The Losses as Implemented}
\label{app:lossimpl}
This section states the auxiliary terms in the form the training code evaluates them.

\paragraph{Whitening and eigensolve.} For each selected molecule the code assembles the full predicted Hamiltonian from the model's diagonal and off-diagonal blocks, permutes it from the backbone's basis ordering into the PySCF ordering, and casts to float64 on the CPU. With the Cholesky factorization $\Sm=\mathbf{L}\mathbf{L}^\top$ it forms $\mathbf{L}^{-1}\Hh\mathbf{L}^{-\top}$, calls a symmetric eigensolver, and back-transforms with $\Ch=\mathbf{L}^{-\top}\mathbf{C}_w$, so that $\Ch^\top\Sm\Ch=\mathbf{I}$ holds to roundoff. Gradients flow through the Cholesky factorization and the eigensolver. The reference spectrum $\epsv^\star,\Cs$ is obtained by the identical code path applied to the reference blocks, which guarantees that prediction and target are compared under the same conventions and the same precision.

\paragraph{Sample average and core suppression.} Writing $\mathcal{R}$ for the cached set of $|\mathcal{R}|=256$ sample points and $\mathrm{div}$ for the geometry-only suppression factor, the density term is
\[
\Lrho=\frac{1}{|\mathcal{R}|}\sum_{\mathbf{r}\in\mathcal{R}}\left(\frac{\hat\rho(\mathbf{r})-\rho^\star(\mathbf{r})}{\mathrm{div}(\mathbf{r})}\right)^{\!2},
\qquad
\hat\rho(\mathbf{r})=2\!\!\sum_{i\le n_{\mathrm{occ}}}\!\!|\hat\psi_i(\mathbf{r})|^2 ,
\]
and the two energy-resolved terms replace the single residual by a mismatch averaged over the energy grid as well,
\[
\mathcal{L}=\frac{1}{|\mathcal{R}|\,K}\sum_{\mathbf{r}\in\mathcal{R}}\sum_{k=1}^{K}
\left(\frac{\hat G(\mathbf{r},E_k)-G^\star(\mathbf{r},E_k)}{\mathrm{div}(\mathbf{r})}\right)^{\!2},
\qquad
G(\mathbf{r},E)=\sum_{i=1}^{n}|\psi_i(\mathbf{r})|^2\,s\!\left(\frac{E-\varepsilon_i}{\sigma}\right),
\]
with $s=\Phi$, the standard normal distribution function, for $\Lcum$ and $s(x)=e^{-x^2/2}$, the unit-height Gaussian, for $\Lldos$. The same $\mathrm{div}$ appears in all three terms, so the core region is suppressed identically and no term is a cusp-fitting objective. The orbital sum runs over all $n$ orbitals, not only those inside the window.

\paragraph{Window anchoring.} The grid is $E_k=\mu^\star-1.5+\left(k-1\right)\cdot1.8/(K-1)$ with $K=64$ and $\sigma=0.02$\,Ha, where $\mu^\star=\tfrac12(\varepsilon^\star_{n_{\mathrm{occ}}}+\varepsilon^\star_{n_{\mathrm{occ}}+1})$ is the \emph{reference} Fermi level. Anchoring the window to the reference rather than to the prediction is what keeps the objective from drifting: a model cannot reduce the loss by moving its own Fermi level out of the window, and the identity of Proposition~\ref{prop:identity} at $E=\mu^\star$ holds against a fixed target.

\paragraph{Batching and cost.} Eight molecules per optimizer step contribute to the auxiliary term, drawn uniformly at random from the batch; the element-wise term uses the whole batch. This subsampling is the only randomness that the auxiliary term adds, and Figure~\ref{fig:appsector}c bounds the resulting per-molecule estimator noise. The auxiliary term has no clipping or rescaling of its own; the gradient-norm clipping of Table~\ref{tab:hparams} applies to the gradient of the total loss.

\paragraph{SCF acceleration protocol.} For Table~\ref{tab:cost} the predicted Hamiltonian of each test molecule is used to build an initial guess: we solve the generalized eigenproblem for $\Hh$ with the reference overlap, form the density matrix from the resulting occupied orbitals, and hand it to PySCF as the starting density for an otherwise unmodified B3LYP/def2-SVP calculation with the default DIIS \citep{pulay1980diis} and the default convergence threshold. The baseline is the same calculation started from \texttt{minao}. Reported ratios are over the same 50 molecules, and the total-time column includes the model forward pass and the eigensolve, so a ratio below one is a net saving after paying for inference.

\section{Additional Results}
\label{app:extra}

\subsection{All metrics on the shifted splits}
\label{app:splits}
Table~\ref{tab:splits} lists every metric for the shifted splits summarized in Figure~\ref{fig:splits}.

\begin{table}[H]
\centering
\caption{Results on the shifted QH9 splits (test sets; energies in \uHa, $\psi$ in \%). Same protocol as Table~\ref{tab:main}. Bold marks the best downstream metric per block.}
\label{tab:splits}
\small
\setlength{\tabcolsep}{2.5pt}
\resizebox{\textwidth}{!}{%
\begin{tabular}{lllrrr?rrrrr}
\toprule
backbone & split & method & $H$ & diag & non-diag & HOMO $\downarrow$ & LUMO $\downarrow$ & gap $\downarrow$ & $\epsv_{\occ}$ $\downarrow$ & $\psi$ $\uparrow$ \\
\midrule
QHNet & stable-ood  & pretrained & 73.53 & 122.32 & 70.78 & 561 & 4632 & 4522 & 740 & 95.54 \\
      &             & + WALoss & 72.70 & 115.00 & 70.30 & 475 & 5511 & 5671 & 393 & 95.04 \\
      &             & + cumulative LDOS (ours) & 71.10 & 110.81 & 68.87 & \textbf{442} & \textbf{1420} & \textbf{1506} & \textbf{290} & \textbf{96.17} \\
\cmidrule(lr){2-11}
      & dynamic-geo & pretrained & 98.4 & 161.4 & 93.6 & 960 & 5561 & 6090 & 840 & 96.66 \\
      &             & + WALoss & 100.0 & 162.2 & 95.3 & 976 & 6950 & 7441 & 597 & 96.06 \\
      &             & + cumulative LDOS (ours) & 95.6 & 148.5 & 91.6 & \textbf{771} & \textbf{1517} & \textbf{2132} & \textbf{400} & \textbf{97.85} \\
\cmidrule(lr){2-11}
      & dynamic-mol & pretrained & 173.5 & 420.1 & 153.7 & 7662 & 68568 & 62262 & 9696 & 79.74 \\
      &             & + WALoss & 174.5 & 419.4 & 154.8 & 8101 & 74057 & 67285 & 9280 & 79.10 \\
      &             & + cumulative LDOS (ours) & 172.3 & 415.9 & 152.8 & \textbf{2494} & \textbf{15105} & \textbf{14356} & \textbf{4842} & \textbf{90.57} \\
\midrule
SPHNet & stable-ood  & pretrained & 30.38 & 38.62 & 29.44 & 171 & 911 & 1004 & 110 & 98.62 \\
       &             & + WALoss & 30.95 & 39.60 & 29.95 & 187 & 605 & 720 & 107 & 98.67 \\
       &             & + cumulative LDOS (ours) & 30.47 & 38.87 & 29.51 & \textbf{156} & \textbf{504} & \textbf{590} & \textbf{105} & \textbf{98.70} \\
\cmidrule(lr){2-11}
       & dynamic-geo & pretrained & 37.77 & 41.27 & 37.23 & 225 & 758 & 928 & 104 & 99.39 \\
       &             & + WALoss & 38.55 & 42.73 & 37.90 & 241 & 642 & 854 & 104 & 99.49 \\
       &             & + cumulative LDOS (ours) & 37.74 & 41.21 & 37.20 & \textbf{203} & \textbf{431} & \textbf{592} & \textbf{98} & \textbf{99.53} \\
\cmidrule(lr){2-11}
       & dynamic-mol & pretrained & 102.32 & 221.59 & 83.38 & 1461 & 13758 & 13433 & 2862 & 93.54 \\
       &             & + WALoss & 102.09 & 221.60 & 83.12 & 1344 & 10632 & 10466 & 2346 & 94.11 \\
       &             & + cumulative LDOS (ours) & 102.28 & 221.43 & 83.36 & \textbf{1306} & \textbf{8528} & \textbf{8362} & \textbf{2068} & \textbf{94.56} \\
\midrule
WANet & stable-ood  & pretrained & 37.05 & 78.99 & 32.28 & 380 & 1259 & 1055 & 160 & 98.55 \\
      &             & + WALoss & 34.42 & 49.26 & 32.72 & 176 & 524 & 577 & 113 & 98.57 \\
      &             & + cumulative LDOS (ours) & 34.09 & 48.39 & 32.46 & \textbf{162} & \textbf{486} & \textbf{549} & \textbf{110} & \textbf{98.59} \\
\cmidrule(lr){2-11}
      & dynamic-geo & pretrained & 41.04 & 46.63 & 40.19 & 269 & 953 & 1126 & 106 & 99.32 \\
      &             & + WALoss & 41.93 & 48.54 & 40.93 & 252 & 675 & 898 & 103 & 99.45 \\
      &             & + cumulative LDOS (ours) & 41.06 & 46.64 & 40.20 & \textbf{217} & \textbf{440} & \textbf{609} & \textbf{97} & \textbf{99.51} \\
\cmidrule(lr){2-11}
      & dynamic-mol & pretrained & 127.51 & 290.69 & 101.52 & 3238 & 26236 & 24126 & 4112 & 89.17 \\
      &             & + WALoss & 127.15 & 290.41 & 101.15 & 3070 & 23111 & 21184 & 3571 & 89.88 \\
      &             & + cumulative LDOS (ours) & 127.46 & 290.44 & 101.51 & \textbf{2020} & \textbf{13756} & \textbf{12936} & \textbf{2186} & \textbf{92.47} \\
\bottomrule
\end{tabular}}
\end{table}

\clearpage
\subsection{Element-wise-only control}
\label{app:honly}
\paragraph{Setting.} Every fine-tuning run in this paper starts from the same pretrained checkpoint and trains for 30k steps with the same optimizer, learning-rate schedule, batch size and data (Table~\ref{tab:hparams}); the runs differ only in the auxiliary loss. The \emph{element-wise-only} run is the control that removes the auxiliary loss altogether: it continues training the pretrained model with its own element-wise loss $\Lelem$ (the mean entry-wise error between $\Hh$ and $\Hs$, exactly the objective used in pretraining) for the same 30k steps, so it simply keeps minimizing the Hamiltonian MAE and nothing else. It isolates the effect of the additional training itself, so that any difference between the WALoss or cumulative-LDOS runs and this control is due to the auxiliary signal and not to the extra steps. Tables~\ref{tab:honly_id} and~\ref{tab:honly_qhnet} repeat the rows of Tables~\ref{tab:main} and~\ref{tab:splits} with this control added, for all three backbones on stable-id and for QHNet on all four splits.

\paragraph{Finding.} Continued element-wise training does what it is asked to do and no more. The element-wise error goes down or stays put, while the eigen-derived quantities do not improve and, on QHNet, deteriorate sharply. On QHNet the Hamiltonian MAE falls on every split (by 4.0\%, 3.6\%, 3.4\% and 0.8\% on stable-id, stable-ood, dynamic-geo and dynamic-mol), while the gap error rises by 53\%, 246\%, 236\% and 19\%, $\psi$ falls on every split, and on dynamic-mol the occupied-orbital error rises as well. On SPHNet every metric changes by less than 1.5\% (the gap error by 0.1\%); on WANet the Hamiltonian MAE is unchanged while the LUMO and gap errors rise by 10\% and 8\%. A matrix that is more accurate entry by entry therefore does not by itself give a better spectrum. The same decoupling has been reported independently: \citet{li2025wanet} found that on larger molecules matrices with tiny relative errors gave unusable total energies, and \citet{kaniselvan2025helm} measured only a weak correlation between matrix-element errors and eigenvalue errors across the test splits of $\nabla^2$DFT. This is Proposition~\ref{prop:realloc} read in reverse: the element-wise loss does not distinguish the sectors of the error, so without a spectral signal the optimizer is free to move error into the sectors that the properties are sensitive to.
\paragraph{Two directions of generalization.} Table~\ref{tab:honly_id} holds the dataset fixed (stable-id) and varies the model: the control behaves the same way on three architectures of very different accuracy, so the effect is a property of the training signal rather than of one network. Table~\ref{tab:honly_qhnet} holds the model fixed (QHNet) and varies the data: the same pattern appears on the random, size-extrapolation and molecular-dynamics splits, so it is not tied to one distribution either. In both tables the cumulative LDOS loss, which shares every setting with the control except the auxiliary term, improves every eigen-derived metric while keeping the Hamiltonian MAE at most 0.3\% above the pretrained value.

\clearpage
\begin{table}[H]
\centering\small
\caption{QH9-stable-id, all three backbones (Table~\ref{tab:main} with the element-wise-only control added; test set, \uHa, $\psi$ in \%). Bold marks the best downstream metric per backbone.}
\label{tab:honly_id}
\setlength{\tabcolsep}{2.5pt}
\resizebox{\textwidth}{!}{%
\begin{tabular}{llrrr?rrrrr}
\toprule
backbone & method & $H$ & diag & non-diag & HOMO $\downarrow$ & LUMO $\downarrow$ & gap $\downarrow$ & $\epsv_{\occ}$ $\downarrow$ & $\psi$ $\uparrow$ \\
\midrule
\multirow{4}{*}{QHNet}
& pretrained & 77.56 & 129.25 & 73.68 & 1041 & 9445 & 9632 & 1206 & 96.28 \\
& + element-wise loss only (control) & 74.44 & 108.11 & 71.90 & 1280 & 14946 & 14769 & 868 & 95.57 \\
& + WALoss & 78.04 & 119.03 & 74.95 & 847 & 5458 & 5835 & 570 & 96.79 \\
& + cumulative LDOS (ours) & 74.81 & 109.54 & 72.19 & \textbf{678} & \textbf{1482} & \textbf{2027} & \textbf{341} & \textbf{97.91} \\
\midrule
\multirow{4}{*}{SPHNet}
& pretrained & 36.60 & 44.79 & 35.38 & 345 & 1309 & 1554 & 380 & 99.06 \\
& + element-wise loss only (control) & 36.55 & 44.67 & 35.35 & 349 & 1311 & 1555 & 381 & 99.05 \\
& + WALoss & 37.28 & 46.03 & 35.98 & 341 & 957 & 1259 & 288 & 99.17 \\
& + cumulative LDOS (ours) & 36.64 & 44.83 & 35.42 & \textbf{292} & \textbf{572} & \textbf{789} & \textbf{137} & \textbf{99.25} \\
\midrule
\multirow{4}{*}{WANet}
& pretrained & 40.07 & 51.28 & 38.42 & 335 & 1454 & 1684 & 288 & 98.98 \\
& + element-wise loss only (control) & 40.08 & 51.25 & 38.43 & 335 & 1603 & 1827 & 292 & 98.95 \\
& + WALoss & 40.99 & 53.26 & 39.19 & 328 & 1025 & 1306 & 157 & 99.13 \\
& + cumulative LDOS (ours) & 40.19 & 51.59 & 38.52 & \textbf{287} & \textbf{573} & \textbf{791} & \textbf{137} & \textbf{99.22} \\
\bottomrule
\end{tabular}}
\end{table}

\begin{table}[H]
\centering\small
\caption{QHNet on all four QH9 splits (the QHNet rows of Tables~\ref{tab:main} and~\ref{tab:splits} with the element-wise-only control added; test sets, \uHa, $\psi$ in \%). Bold marks the best downstream metric per split.}
\label{tab:honly_qhnet}
\setlength{\tabcolsep}{2.5pt}
\resizebox{\textwidth}{!}{%
\begin{tabular}{llrrr?rrrrr}
\toprule
split & method & $H$ & diag & non-diag & HOMO $\downarrow$ & LUMO $\downarrow$ & gap $\downarrow$ & $\epsv_{\occ}$ $\downarrow$ & $\psi$ $\uparrow$ \\
\midrule
\multirow{4}{*}{stable-id}
& pretrained & 77.56 & 129.25 & 73.68 & 1041 & 9445 & 9632 & 1206 & 96.28 \\
& + element-wise loss only (control) & 74.44 & 108.11 & 71.90 & 1280 & 14946 & 14769 & 868 & 95.57 \\
& + WALoss & 78.04 & 119.03 & 74.95 & 847 & 5458 & 5835 & 570 & 96.79 \\
& + cumulative LDOS (ours) & 74.81 & 109.54 & 72.19 & \textbf{678} & \textbf{1482} & \textbf{2027} & \textbf{341} & \textbf{97.91} \\
\midrule
\multirow{4}{*}{stable-ood}
& pretrained & 73.53 & 122.32 & 70.78 & 561 & 4632 & 4522 & 740 & 95.54 \\
& + element-wise loss only (control) & 70.86 & 109.25 & 68.70 & 783 & 15801 & 15638 & 582 & 94.35 \\
& + WALoss & 72.70 & 115.00 & 70.30 & 475 & 5511 & 5671 & 393 & 95.04 \\
& + cumulative LDOS (ours) & 71.10 & 110.81 & 68.87 & \textbf{442} & \textbf{1420} & \textbf{1506} & \textbf{290} & \textbf{96.17} \\
\midrule
\multirow{4}{*}{dynamic-geo}
& pretrained & 98.4 & 161.4 & 93.6 & 960 & 5561 & 6090 & 840 & 96.66 \\
& + element-wise loss only (control) & 95.1 & 146.4 & 91.3 & 1503 & 20655 & 20468 & 770 & 94.51 \\
& + WALoss & 100.0 & 162.2 & 95.3 & 976 & 6950 & 7441 & 597 & 96.06 \\
& + cumulative LDOS (ours) & 95.6 & 148.5 & 91.6 & \textbf{771} & \textbf{1517} & \textbf{2132} & \textbf{400} & \textbf{97.85} \\
\midrule
\multirow{4}{*}{dynamic-mol}
& pretrained & 173.5 & 420.1 & 153.7 & 7662 & 68568 & 62262 & 9696 & 79.74 \\
& + element-wise loss only (control) & 172.1 & 415.8 & 152.5 & 9145 & 82180 & 74214 & 11343 & 76.80 \\
& + WALoss & 174.5 & 419.4 & 154.8 & 8101 & 74057 & 67285 & 9280 & 79.10 \\
& + cumulative LDOS (ours) & 172.3 & 415.9 & 152.8 & \textbf{2494} & \textbf{15105} & \textbf{14356} & \textbf{4842} & \textbf{90.57} \\
\bottomrule
\end{tabular}}
\end{table}

\clearpage
\subsection{Weight sensitivity on QHNet}
\label{app:dose}
\paragraph{WALoss as used in this paper.} Throughout the paper, WALoss means the full form of \citet{li2025wanet}: the prediction is expressed in the basis of reference orbitals and every entry of $\Cs{}^\top\Hh\Cs-\diag(\epsv^\star)$ is penalized. On SPHNet and WANet, the loss is the mean absolute plus the mean squared entry of this matrix, with weight $0.05$. On QHNet, it is the mean absolute entry, evaluated on eight molecules per batch like our own loss, and we report two weights: $w=0.03$, at which the WALoss term is about as large as the element-wise term, and $w=0.1$, at which it dominates; both are in Table~\ref{tab:dose} and give similar gap errors (6072 and 5835\,\uHa).

\begin{table}[H]
\centering\small
\caption{QHNet on QH9-stable-id: all metrics against the auxiliary weight $w$ (test set, \uHa). WALoss is the full form used throughout the paper; the weight of the cumulative loss, $w=30$, is selected on the validation split.}
\label{tab:dose}
\setlength{\tabcolsep}{3pt}
\resizebox{\textwidth}{!}{%
\begin{tabular}{lrrr?rrrrr}
\toprule
method & $H$ & diag & non-diag & HOMO & LUMO & gap & $\epsv_{\occ}$ & $\psi$ \\
\midrule
pretrained & 77.56 & 129.25 & 73.68 & 1041 & 9445 & 9632 & 1206 & 96.28 \\
element-wise loss only (control) & 74.44 & 108.11 & 71.90 & 1280 & 14946 & 14769 & 868 & 95.57 \\
\midrule
WALoss $w{=}0.03$ & 76.01 & 112.51 & 73.26 & 823 & 5728 & 6072 & 593 & 96.81 \\
WALoss $w{=}0.1$ & 78.04 & 119.03 & 74.95 & 847 & 5458 & 5835 & 570 & 96.79 \\
\midrule
density only $w{=}1$ & 77.72 & 126.93 & 74.02 & 780 & 1923 & 2376 & 571 & 97.44 \\
density only $w{=}10$ & 79.01 & 138.44 & 74.55 & 802 & 2500 & 2896 & 373 & 97.43 \\
density only $w{=}100$ & 76.82 & 118.70 & 73.65 & 719 & 1575 & 2081 & 356 & 97.87 \\
cumulative LDOS $w{=}10$ & 74.88 & 109.66 & 72.26 & 675 & 1693 & 2231 & 348 & 97.80 \\
cumulative LDOS $w{=}30$ & 74.81 & 109.54 & 72.19 & 678 & 1482 & 2027 & 341 & 97.91 \\
cumulative LDOS $w{=}100$ & 74.85 & 109.90 & 72.21 & 655 & 1300 & 1811 & 336 & 97.98 \\
\bottomrule
\end{tabular}}
\end{table}

\begin{figure}[H]
\centering
\includegraphics[width=\textwidth]{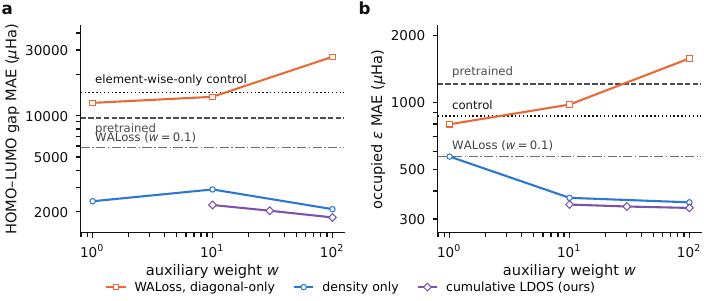}
\caption{Weight sensitivity on QHNet (entries of Tables~\ref{tab:dose} and~\ref{tab:diagwa}). \textbf{(a)} Gap error and \textbf{(b)} occupied-orbital energy error against the auxiliary weight $w$, with the pretrained checkpoint, the element-wise-only control and WALoss ($w=0.1$) as horizontal references. The diagonal-only WALoss gets worse as its weight grows; the density-only and cumulative losses improve or stay flat, and the cumulative loss is the lowest at every weight.}
\label{fig:appdose}
\end{figure}

Figure~\ref{fig:appdose} plots the density-only and cumulative rows of Table~\ref{tab:dose} and the diagonal-only rows of Table~\ref{tab:diagwa} against $w$, with the pretrained model, the control and WALoss ($w=0.1$) as horizontal references. Across all tested weights, the density-only and cumulative losses give 2--3 times lower gap error than WALoss, so the ranking does not depend on the choice of $w$. The cumulative loss responds only mildly to its weight (gap error 2231, 2027 and 1811\,\uHa\ at $w=10$, 30 and 100) and its element-wise error stays essentially constant: at the selected weight, $w=30$, its element-wise error is within 0.6\% of the lowest value among all runs, that of the control, while its gap error is 7.3 times lower.

\clearpage
\paragraph{The diagonal-only form.} \citet{li2025wanet} also propose a second form of WALoss that keeps only the diagonal of the projected matrix, $\sum_i w_i\big(\mathbf{c}_i^{\star\top}\Hh\,\mathbf{c}^\star_i-\varepsilon^\star_i\big)^2$, with large weights $w_i$ on the occupied orbitals and the LUMO. Each term is the first-order estimate of one orbital energy (Proposition~\ref{prop:sectors}a), so this form sees sector A only, and only to first order. We implemented it on QHNet ($w_i=1$ on the occupied orbitals and the LUMO, $0.01$ elsewhere) to test the prediction of Sec.~\ref{sec:analysis} that such an objective can be lowered by rotating orbitals instead of correcting their energies. Table~\ref{tab:diagwa} confirms it: as the weight grows, the gap and occupied-orbital errors get monotonically worse, from 12{,}373 to 26{,}813\,\uHa\ and from 797 to 1573\,\uHa, while $\psi$ falls.

\begin{table}[H]
\centering\small
\caption{QHNet on QH9-stable-id with the diagonal-only form of WALoss at three weights (test set, \uHa).}
\label{tab:diagwa}
\setlength{\tabcolsep}{3pt}
\resizebox{\textwidth}{!}{%
\begin{tabular}{lrrr?rrrrr}
\toprule
method & $H$ & diag & non-diag & HOMO & LUMO & gap & $\epsv_{\occ}$ & $\psi$ \\
\midrule
pretrained & 77.56 & 129.25 & 73.68 & 1041 & 9445 & 9632 & 1206 & 96.28 \\
diagonal-only WALoss $w{=}1$ & 74.48 & 108.29 & 71.93 & 1130 & 12504 & 12373 & 797 & 95.88 \\
diagonal-only WALoss $w{=}10$ & 75.20 & 110.03 & 72.58 & 1090 & 14208 & 13701 & 977 & 95.59 \\
diagonal-only WALoss $w{=}100$ & 77.84 & 117.44 & 74.86 & 1862 & 28282 & 26813 & 1573 & 93.66 \\
\bottomrule
\end{tabular}}
\end{table}

\clearpage
\subsection{Peak-shaped LDOS variants}
\label{app:gauss}
The peak-shaped loss $\Lldos$ of Sec.~\ref{sec:loss} replaces the cumulative step by a unit-height Gaussian; everything else (window, $\sigma$, sample points) is shared with the cumulative loss. On SPHNet and WANet at $w=150$ it reaches gap errors of 805 and 813\,\uHa\ (HOMO 281 and 270, LUMO 588 and 613, $\epsv_{\occ}$ 235 and 140, $\psi$ 99.22 and 99.19\%), close to the cumulative loss at $w=450$ (789 and 791\,\uHa), so on these two backbones the two forms perform similarly. On QHNet, whose initial orbital-energy errors are the largest, the peak-shaped loss is unreliable: two runs with identical settings end at gap errors of 6917 and 1994\,\uHa, and in the first run the occupied-orbital error doubles relative to the pretrained model. The three cumulative runs of Table~\ref{tab:dose}, by contrast, vary smoothly with $w$ (gap error 2231, 2027 and 1811\,\uHa). Per-orbital errors (Table~\ref{tab:orbitals}) show why: the deep occupied orbitals, whose errors are largest, drift away from their targets (for HOMO$-23$, the 23rd orbital below the HOMO, the error grows from 1{,}049 to 20{,}979\,\uHa), whereas the cumulative loss, with the same window, $\sigma$ and weight, shows no drift and gives the best $\epsv_{\occ}$ and $\psi$ of all variants at the same element-wise error.

The mechanism is Theorem~\ref{thm:props}(iv): once an orbital's energy error exceeds a few $\sigma$, the predicted and reference peaks no longer overlap, the mismatch saturates and its gradient vanishes (Figure~\ref{fig:sector}b,c), so the orbital is no longer pulled back; the step of the cumulative form keeps a constant restoring force at any distance. This is the failure mode of overlap-based divergences between distributions with disjoint supports, whose gradients vanish while a transport (Wasserstein) distance keeps a usable one \citep{arjovsky2017wgan}; the cumulative form compares the local spectra through their distribution functions, as a transport distance does, and tends to their Cram\'er--von Mises distance as $\sigma\to0$ (Proposition~\ref{prop:saturation}). The deep occupied orbitals, whose initial errors on QHNet are the largest, are the first to be released.
\begin{table}[H]
\centering\small
\caption{Per-orbital energy error on QHNet (1{,}000 test molecules, \uHa) for the peak-shaped and the cumulative loss at the same weight ($w=30$), window and $\sigma$. The peak-shaped loss corrects the virtual orbitals inside the window but lets the deep occupied orbitals drift; the cumulative loss improves every orbital.}
\label{tab:orbitals}
\resizebox{\textwidth}{!}{%
\begin{tabular}{lrrrrrrrr}
\toprule
loss & LUMO$+2$ & LUMO$+1$ & LUMO & HOMO & HOMO$-5$ & HOMO$-19$ & HOMO$-21$ & HOMO$-23$ \\
\midrule
pretrained & 3455 & 3838 & 9593 & 1005 & 403 & 635 & 1029 & 1049 \\
peak LDOS & 2287 & 1698 & 7685 & 1053 & 593 & 5955 & 11364 & 20979 \\
cumulative LDOS & \textbf{1032} & \textbf{994} & \textbf{1567} & \textbf{690} & \textbf{235} & \textbf{195} & \textbf{251} & \textbf{367} \\
\bottomrule
\end{tabular}}
\end{table}
\clearpage
\subsection{SCF acceleration}
\label{app:scf}
A predicted Hamiltonian can also initialize a conventional DFT calculation (the protocol is in Appendix~\ref{app:lossimpl}). On 50 stable-id test molecules, every model in Table~\ref{tab:cost} saves about a third of the iterations and 13--15\% of the total time relative to the \texttt{minao} guess, and the models differ little (0.671--0.674 in iterations, 0.80--0.82 in SCF time). The loss improves the eigen-properties without changing what the model offers as an SCF starting point.

\begin{table}[H]
\centering\small
\caption{SCF acceleration on 50 QH9-stable-id test molecules: ratio of SCF iterations and of wall time relative to the \texttt{minao} initial guess (lower is better; total time includes model inference).}
\label{tab:cost}
\begin{tabular}{lrrr}
\toprule
method & SCF iterations & SCF time & total time \\
\midrule
pretrained & 0.6732 & 0.8115 & 0.8619 \\
element-wise loss only (control) & 0.6735 & 0.8139 & 0.8655 \\
density only ($w=100$) & 0.6710 & 0.8171 & 0.8682 \\
cumulative LDOS ($w=100$) & 0.6713 & 0.8028 & 0.8505 \\
\bottomrule
\end{tabular}
\end{table}

\subsection{Training cost}
\label{app:cost}
With the cached orbital values, evaluating $\Nre$ costs one eigendecomposition and two small matrix products per molecule; QH9 matrices are at most $226\times226$. On QHNet the auxiliary term adds 0.72\,s to each training step and no GPU memory, because the eigendecompositions run on the CPU (Table~\ref{tab:traincost}); over the 30k fine-tuning steps this amounts to about 6 hours, roughly 5\% of the cost of pretraining QHNet for 300k steps. The loss is used only in training, so inference costs exactly what it costs for the backbone.
\begin{table}[H]
\centering\small
\caption{Training cost of the auxiliary losses on QHNet (QH9-stable-id), measured on one NVIDIA L40S with the same code, data order and batch size (32). Seconds per step are averaged over steps 50--300 without validation; memory is the peak reserved CUDA memory. The eigendecompositions run on the CPU in double precision.}
\label{tab:traincost}
\begin{tabular}{lrrr}
\toprule
auxiliary loss & seconds/step & overhead & peak GPU memory (GB) \\
\midrule
none (element-wise only) & 1.47 & -- & 31.2 \\
density only & 1.92 & $+0.45$\,s ($+30\%$) & 31.2 \\
peak LDOS & 2.28 & $+0.81$\,s ($+55\%$) & 31.2 \\
cumulative LDOS & 2.19 & $+0.72$\,s ($+49\%$) & 31.2 \\
\bottomrule
\end{tabular}
\end{table}

\section{Numerical Verification}
\label{app:verify}
\begin{figure}[H]
\centering
\includegraphics[width=\textwidth]{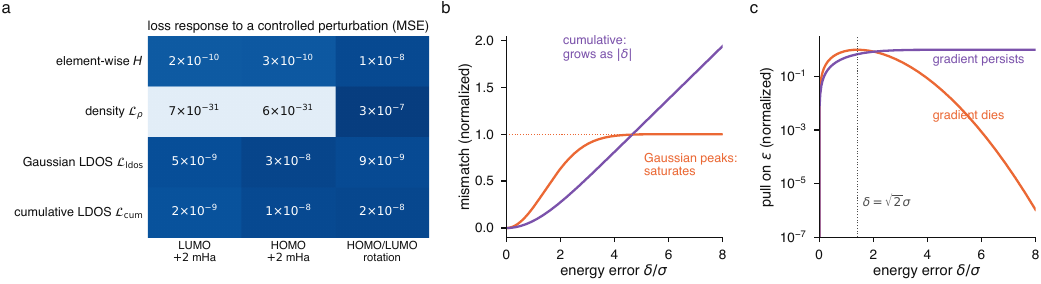}
\caption{\textbf{(a)} Response of each loss to controlled perturbations of the reference Hamiltonian of one QH9 molecule ($\#12345$): shifting the LUMO or HOMO energy by 2\,mHa with all orbitals fixed, or rotating the HOMO and LUMO into each other with all energies fixed. Density matching does not respond to energy shifts at all (values at the level of floating-point round-off); both LDOS losses respond to both kinds of error. \textbf{(b)} Mismatch between a displaced and a reference orbital as a function of the energy error $\delta$, for a Gaussian peak and for a cumulative step (closed forms of Theorem~\ref{thm:props}(iv)): the peak mismatch saturates, the step mismatch keeps growing. \textbf{(c)} The corresponding gradient with respect to $\delta$, normalized by its maximum: the peak gradient vanishes beyond a few $\sigma$, whereas the step gradient approaches a constant.}
\label{fig:sector}
\end{figure}

Every analytical statement in Sections~\ref{sec:background} and \ref{sec:method} was checked numerically on reference data. Table~\ref{tab:verify} collects the checks and their outcomes. The Fermi-level identity holds to the single precision in which the orbital values are cached: the residual scales with the float32 unit round-off. The two perturbations used in Figure~\ref{fig:sector}a are exact by construction: the energy-only shift moves the intended orbital energy by $2.000$\,mHa with an eigenvector overlap deviation at the level of double-precision round-off, and the shape rotation leaves every eigenvalue fixed to $2.4\times10^{-13}$\,Ha, so the sector test cleanly separates energy errors from shape errors.

The lower block of Table~\ref{tab:verify} checks Theorem~\ref{thm:complete} and the statements around it on the reference Hamiltonians of three QH9-stable molecules, $\#18$ ($n=81$), $\#12345$ ($n=162$) and $\#23470$ ($n=151$), with orbital values evaluated on the level-0 quadrature grid. For injectivity, we form the first-order response of $\Nre$ to every entry of a symmetric $\Dl$, one column per entry, normalize the columns, and report the smallest singular value relative to the largest. For recovery, we start from random symmetric errors with $\|\Dl\|_2=50$\,mHa on molecule $\#18$ and minimize the mismatch in $\Nre$ over the full energy range, or in $\rho$, over $\Dl$. Minimizing the mismatch in $\Nre$ recovers every orbital energy to $0.03$\,mHa and shrinks the error among orbitals inside the window about 30-fold; the residual, 20--30\% of the initial $\|\Dl\|_F$, lies mostly in couplings between core and virtual orbitals, the directions to which $\Nre$ is least sensitive. Minimizing the density mismatch instead leaves the orbital energies wrong by $10$\,mHa. The supplementary code contains the scripts for these checks (\texttt{tests/check.py} and \texttt{tests/verify\_theory.py}).

\clearpage
\begingroup
\footnotesize
\setlength{\tabcolsep}{3pt}
\setlength{\LTcapwidth}{\textwidth}
\renewcommand{\arraystretch}{1.2}
\begin{longtable}{@{}>{\raggedright\arraybackslash}p{0.21\textwidth}
                       >{\raggedright\arraybackslash}p{0.34\textwidth}
                       >{\raggedright\arraybackslash}p{0.28\textwidth}
                       >{\raggedright\arraybackslash}p{\dimexpr0.17\textwidth-6\tabcolsep\relax}@{}}
\caption{Numerical checks. Molecule $\#12345$ and $\#23470$ are the two QH9-stable geometries used in the appendix figures; the lower block also uses molecule $\#18$. ``Machine zero'' means the response is at or below the double-precision noise floor of the quantity being differenced.}
\label{tab:verify}\\
\toprule
statement & what is checked & result & reference \\
\midrule
\endfirsthead
\toprule
statement & what is checked & result & reference \\
\midrule
\endhead
\bottomrule
\endfoot
generalized eigenbasis & $\|\Ch^\top\Sm\Ch-\mathbf{I}\|_\infty$ & $4.8\times10^{-14}$ & -- \\
cache consistency & recomputed vs.\ released $\Sm$ & $2.2\times10^{-7}$ & App.~\ref{app:data} \\
cache consistency & $\int\rho^\star\,dV$ vs.\ $N_e=60$ & $59.978$ & App.~\ref{app:data} \\
Prop.~\ref{prop:identity} & $\Nre(\mathbf{r},\mu)$ vs.\ $\rho^\star(\mathbf{r})/2$, all 256 points & $6.4\times10^{-8}$ rel.\ (float32 cache) & Fig.~\ref{fig:appspectral}c \\
Prop.~\ref{prop:saturation} & closed forms $g(\delta)$, $c(\delta)$ vs.\ quadrature & $9.3\times10^{-16}$ rel. & Fig.~\ref{fig:sector}b \\
energy-only perturbation & induced $\Delta\varepsilon_i$ & $2.000$\,mHa (exact) & Fig.~\ref{fig:sector}a \\
energy-only perturbation & eigenvector overlap deviation & $\le4.4\times10^{-16}$ & Fig.~\ref{fig:sector}a \\
shape-only perturbation & $\max_i|\Delta\varepsilon_i|$ & $2.4\times10^{-13}$\,Ha & Fig.~\ref{fig:sector}a \\
Prop.~\ref{prop:blind} & $\Lrho$ response to a $2$\,mHa energy shift & $6.2$--$6.6\times10^{-31}$ (machine zero) & Fig.~\ref{fig:sector}a \\
Prop.~\ref{prop:sectors}(c) & $\Lrho$ response to a $0.05$\,rad rotation & $2.7\times10^{-7}$ & Fig.~\ref{fig:sector}a \\
Gaussian saturation & $|g'|$ at $\delta=4\sigma$, $6\sigma$, rel.\ to its peak & $8.6\times10^{-2}$, $8.7\times10^{-4}$ & Fig.~\ref{fig:sector}c \\
\midrule
Prop.~\ref{prop:sectors}(a,b), first-order $\delta\Nre$ & exact minus first-order change, perturbation scale $t=10^{-5},10^{-6},10^{-7}$ & rel.\ error $\propto t$ ($3.8\times10^{-4}\to3.8\times10^{-6}$) & Thm.~\ref{thm:complete}(i) \\
Prop.~\ref{prop:sectors}(c), degenerate levels & same for $\delta\rho$, two occupied and two virtual levels degenerate & rel.\ error $\propto t$ ($6.5\times10^{-5}\to7.3\times10^{-7}$) & App.~\ref{app:proofs} \\
Prop.~\ref{prop:identity} at the grid energies $\mu\pm0.014$\,Ha & $\max|\Nre-\rho/2|/\max(\rho/2)$ on the grid; ratio to the bound & $\le8.5\times10^{-8}$; $\le0.22$ & App.~\ref{app:proofs} \\
Thm.~\ref{thm:complete}(i), full energy range & smallest relative singular value of the first-order response & $\ge6.1\times10^{-6}$, no null direction & App.~\ref{app:proofs} \\
Thm.~\ref{thm:complete}(i), loss window ($K=64$) & directions with an orbital inside the window / with both outside & all resolved, $\ge2.5\times10^{-7}$ / below round-off & App.~\ref{app:proofs} \\
Thm.~\ref{thm:complete}(ii), recovery & max orbital-energy error after minimizing $\Lcum$ (or $\Lrho$), three random starts & $0.03$\,mHa ($\Lrho$: $10$\,mHa) & App.~\ref{app:proofs} \\
Davis--Kahan bound & $\|\sin\Theta\|_F$ over the bound, 600 random $\Dl$ with $\|\Dl\|_2<g/2$ & $\le0.22$ & App.~\ref{app:proofs} \\
restoring force on the $K=64$ grid & $|c'-\erf(\delta/2\sigma)|$, $|\delta|\le0.1$\,Ha, levels $\ge0.18$\,Ha inside & $\le7.2\times10^{-3}$ & Prop.~\ref{prop:saturation} \\
gradient of $\Lcum$ & Daleckii--Krein form vs.\ central differences & $1.9\times10^{-9}$ rel. & App.~\ref{app:proofs} \\
two $p$ shells on one atom & $\max|p_{1x}p_{2y}-p_{1y}p_{2x}|/\max|p_{1x}p_{2y}|$ & $2.9\times10^{-16}$ (identically zero) & App.~\ref{app:proofs} \\
\end{longtable}
\endgroup

\section{Limitations and Future Work}
\label{app:limitations}
The analysis is exact within its assumptions: first-order perturbation theory in the basis of reference orbitals, with non-degenerate reference energies for the statements about orbital energies and shapes and an open HOMO--LUMO gap for the statement about the density. Under these assumptions the sector picture is a change of basis rather than an approximation. What the analysis does not do is predict how a given network moves along the sectors during training, so the sector picture is best read as a diagnostic rather than a guarantee.

The loss has been tested where Hamiltonian prediction is currently benchmarked: on QH9 (B3LYP/def2-SVP), with three architectures that span the development of the field, on all four official splits, and against the strongest published physics-informed loss. Its ingredients, the predicted Hamiltonian, the overlap matrix and orbital values at a few hundred points, exist for any basis set, functional or system, so the same construction should carry over to larger molecules, other levels of theory and periodic systems, where the LDOS is resolved per $k$-point and is the quantity that scanning tunnelling microscopy measures; these are the settings in which we plan to test it next. Two design choices should be noted: the energy window covers the valence and low virtual orbitals, the right choice for ground-state properties, and would be widened for excited-state targets; and the smoothing width $\sigma$ was fixed rather than tuned.

The cost is one eigendecomposition per sampled molecule and training step, the same operation that reads any property from a predicted Hamiltonian; on QH9 it adds 0.72\,s per step on QHNet (Appendix~\ref{app:cost}), and inference is untouched. At the scale of thousands of basis functions the dense solver would give way to the iterative and fragment-based solvers that DFT codes already use.

Two directions follow from the analysis itself. The cumulative LDOS can be used from the start of training rather than as a fine-tuning signal, where its non-vanishing gradients should matter most while the spectrum is still far from the reference. And the sector representation can serve as a diagnostic in its own right: computed on a validation set, it tells which part of the residual a model still carries and therefore which training signal will help, turning the choice of loss from trial and error into a measurement.

\section{Implementation Details and Protocol}
\label{app:impl}
\paragraph{Sample points and core suppression.} Each molecule's sample set $\mathcal{R}$ is a fixed random subset of 256 points of the atom-centred radial--Lebedev quadrature grid of PySCF at level 0, cached once. The core-suppression factor $\mathrm{div}(\mathbf{r})$ is the element-wise fit of \citet{ushenin2025lagnet}, which depends only on the geometry.

\paragraph{Eigensolve.} For each of the eight molecules per step we form $\mathbf{L}^{-1}\Hh\mathbf{L}^{-\top}$ with $\Sm=\mathbf{L}\mathbf{L}^\top$, call a symmetric float64 eigensolver on the CPU, and back-transform; gradients flow through the eigensolver.

\paragraph{Energy grids and weights.} $K=64$ points on $[\mu-1.5,\mu+0.3]$\,Ha with $\sigma=0.02$\,Ha for all backbones. The loss weights are listed in Table~\ref{tab:hparams}. The sum in \Eqref{eq:cumldos} runs over all orbitals, so orbitals below the window contribute a constant step and $\Nre(\mathbf{r},E)=\rho/2$ holds at every energy inside the gap, up to the bound of Proposition~\ref{prop:identity}.

\paragraph{Evaluation.} Element-wise MAE is basis-permutation invariant. Eigen-derived metrics are computed from the generalized eigenproblem of $\Hh$ with the DFT overlap $\Sm$. All comparisons are made within a backbone using the backbone's own evaluator and matched settings. For SPHNet and WANet, whose evaluator runs on the GPU, TF32 matrix multiplication is disabled and the generalized eigenproblem is solved in float64, matching the CPU float64 evaluation of QHNet.

\paragraph{Optimization.} Table~\ref{tab:hparams} lists the settings. Every fine-tuning run of a given backbone starts from the same released checkpoint and differs only in the auxiliary term and its weight; the optimizer, learning rate, schedule, batch size, gradient clipping, step budget and data split are held fixed across runs.

\begin{table}[H]
\centering\small
\caption{Fine-tuning settings. Every method of a given backbone shares all settings except the auxiliary loss and its weight; unlisted settings follow each backbone's released training recipe (``native''). The cumulative LDOS weight is selected on the validation split; Table~\ref{tab:dose} reports its test-set sensitivity on QHNet.}
\label{tab:hparams}
\footnotesize
\setlength{\tabcolsep}{4pt}
\begin{tabular}{lll}
\toprule
 & QHNet & SPHNet / WANet \\
\midrule
optimizer & AdamW, $\beta=(0.99,0.999)$ & native \\
learning rate & $1\times10^{-5}$ & $1\times10^{-5}$ \\
warm-up steps & 1{,}000 & native \\
fine-tuning steps & 30{,}000 & 30{,}000 \\
batch size & 32 & 64 \\
gradient clipping & norm 5.0 & norm 5.0 \\
\midrule
molecules per auxiliary step & 8 & 8 \\
energy grid & $K=64$, $\sigma=0.02$\,Ha & $K=64$, $\sigma=0.02$\,Ha \\
cumulative LDOS weight $w$ & 30 (stable-id), 100 (other splits) & 450 \\
WALoss weight & 0.1 & 0.05 \\
\bottomrule
\end{tabular}
\end{table}

\paragraph{Verification scripts.} A sensitivity script perturbs the reference Hamiltonian of one molecule by an eigenvalue-only shift $\Hs+d\,\Sm\mathbf{c}_i\mathbf{c}_i^\top\Sm$ (which moves $\varepsilon_i$ by $d$ and no eigenvector) and by a rotation in the HOMO/LUMO plane, and evaluates each loss (Figure~\ref{fig:sector}a). A second script checks the identity of Prop.~\ref{prop:identity}, the closed forms of Prop.~\ref{prop:saturation} against numerical quadrature, and the sector sensitivities of both LDOS forms. Appendix~\ref{app:verify} reports the outcome of every check.

\end{document}

%% file: math_commands.tex
\usepackage{amsmath,amsfonts,bm}

\def\eqref#1{equation~\ref{#1}}
\def\Eqref#1{Equation~\ref{#1}}

\def\1{\bm{1}}

\DeclareMathAlphabet{\mathsfit}{\encodingdefault}{\sfdefault}{m}{sl}
\SetMathAlphabet{\mathsfit}{bold}{\encodingdefault}{\sfdefault}{bx}{n}



%% file: app-theory.tex
\section{Proofs and Additional Theory}
\label{app:proofs}

Throughout, $\Pm=2\Cs_{\occ}\Cs_{\occ}{}^\top$ denotes the density matrix, so that $\rho(\mathbf{r})=\boldsymbol{\phi}(\mathbf{r})^\top\Pm\,\boldsymbol{\phi}(\mathbf{r})$ with $\boldsymbol{\phi}(\mathbf{r})=(\phi_1(\mathbf{r}),\dots,\phi_n(\mathbf{r}))^\top$; hats denote the same quantities computed from $\Hh$. We write $g:=\varepsilon^\star_{n_o+1}-\varepsilon^\star_{n_o}$ for the reference HOMO--LUMO gap and $\mathcal{U}:=\spn\{\phi_\mu\}_{\mu=1}^{n}$ for the functions representable in the basis, which contains every predicted and reference orbital. Gaussian-type basis functions are real-analytic on $\mathbb{R}^3$ and linearly independent, so every nonzero $u\in\mathcal{U}$ is a nonzero real-analytic function. Two consequences are used below: a nonzero $u\in\mathcal{U}$ does not vanish identically on any nonempty open set, and if $u,v\in\mathcal{U}$ satisfy $uv=0$ on $\mathbb{R}^3$, then $u=0$ or $v=0$ (if $u\ne0$, $v$ vanishes on the dense open set where $u\ne0$, hence everywhere).

\subsection{The error in the orbital basis}
\emph{Whitening} a matrix $\mathbf{M}$ means expressing it in an orthonormal basis of the span of the atomic orbitals, $\mathbf{M}\mapsto\Sm^{-1/2}\mathbf{M}\,\Sm^{-1/2}$ (L\"owdin orthogonalization). The error $\Dl$ of \Eqref{eq:Delta} is the whitened error rotated into the reference eigenbasis, which gives the following identities.

\begin{lemma}[Whitening]
\label{lem:whitening}
$\|\Dl\|_F=\|\Sm^{-1/2}\dH\,\Sm^{-1/2}\|_F\le \|\Sm^{-1}\|_2\,\|\dH\|_F=\tfrac{\kappa(\Sm)}{\|\Sm\|_2}\|\dH\|_F$. Moreover $\Cs{}^\top\Hh\Cs-\diag(\epsv^\star)=\Dl$, so the full WALoss of \citet{li2025wanet}, an average of the absolute or squared entries of this matrix, weights every entry of $\Dl$ alike (its mean-squared form is $\|\Dl\|_F^2/n^2$), and the diagonal-only form equals $\sum_i w_i d_i^2$.
\end{lemma}
\begin{proof}
Write $\Cs=\Sm^{-1/2}\mathbf{Q}$ with $\mathbf{Q}$ orthogonal (from $\Cs{}^\top\Sm\Cs=\mathbf{I}$). Then $\Dl=\mathbf{Q}^\top\Sm^{-1/2}\dH\Sm^{-1/2}\mathbf{Q}$, and the Frobenius norm is orthogonally invariant. The bound follows from submultiplicativity and $\|\Sm^{-1/2}\|_2^2=\|\Sm^{-1}\|_2$. The identity $\Cs{}^\top\Hh\Cs-\diag(\epsv^\star)=\Dl$ is \Eqref{eq:Delta}, because $\Cs{}^\top\Hs\Cs=\diag(\epsv^\star)$ by \Eqref{eq:gep}, and the diagonal of $\Dl$ is $(d_i)$.
\end{proof}
The inequality is the core of the perturbation theorem of \citet{li2025wanet}. On QH9 it is loose: Appendix~\ref{app:realresidual} measures a median amplification $\|\Dl\|_F/\|\dH\|_F$ of 3.0 against a median bound of 1878, so the location of the error, rather than its size, is what determines its effect.
Table~\ref{tab:coverage} applies Propositions~\ref{prop:sectors} and~\ref{prop:blind} to the losses of Section~\ref{sec:background}; Figure~\ref{fig:teaser}e shows the same comparison.

\begin{table}[H]
\centering\small
\caption{Which sectors of the error each loss responds to (first order), how it weights them, and whether it uses the eigenpairs of the prediction. The element-wise loss is uniform up to the overlap metric (Lemma~\ref{lem:whitening}).}
\label{tab:coverage}
\resizebox{\textwidth}{!}{%
\begin{tabular}{lccccc}
\toprule
loss & energies $d_i$ & $\Dl_{\occ\vir}$ (density) & $\Dl_{\occ\occ},\Dl_{\vir\vir}$ & weighting & eigenpairs of $\Hh$ \\
\midrule
element-wise $\Lelem$ & $\checkmark$ & $\checkmark$ & $\checkmark$ & uniform & -- \\
WALoss, full \citep{li2025wanet} & $\checkmark$ & $\checkmark$ & $\checkmark$ & uniform & -- \\
WALoss, diagonal-only \citep{li2025wanet} & first order & -- & -- & occupied orbitals and LUMO & -- \\
density matching $\Lrho$ & -- & $\checkmark$ & -- & frontier, $1/(\varepsilon_a-\varepsilon_i)$ & $\checkmark$ \\
cumulative LDOS $\Lcum$ (ours) & $\checkmark$ & $\checkmark$ & $\checkmark$ & frontier and energy window & $\checkmark$ \\
\bottomrule
\end{tabular}}
\end{table}

\clearpage
\subsection{Perturbation theory for each property}
Besides Proposition~\ref{prop:sectors}, the main text uses two second-order facts.

\begin{proposition}[Second-order corrections]
\label{prop:second}
Under the assumptions of Proposition~\ref{prop:sectors}:
\begin{enumerate}[label=(\alph*),start=4]
\item $\hat\varepsilon_i-\varepsilon^\star_i=d_i+\sum_{j\ne i}\frac{\Delta_{ij}^2}{\varepsilon^\star_i-\varepsilon^\star_j}+O(\|\Dl\|^3)$;
\item the Kohn--Sham total energy is stationary in $\Pm$ at the ground state, so its error is second order in $\Dl_{\occ\vir}$ and has no first-order dependence on any sector.
\end{enumerate}
\end{proposition}
Item (d) is the channel through which off-diagonal errors reach the orbital energies: a coupling $\Delta_{ij}$ between two orbitals close in energy shifts both with weight $1/(\varepsilon^\star_i-\varepsilon^\star_j)$, which is largest across the HOMO--LUMO gap. It is also why a loss that sees only the diagonal, and only to first order, such as the diagonal-only WALoss, can be lowered while the true orbital energies stay wrong. Item (e) explains why total-energy errors are a late indicator of the problem \citep{li2025wanet}: the energy is the least sensitive probe of the occupied subspace and the density the most sensitive one.

\begin{remark}[What the sectors mean for the metrics]
\label{rem:sectors}
To first order, the gap error is $d_{\mathrm{LUMO}}-d_{\mathrm{HOMO}}$, two entries of sector A out of $n(n+1)/2$; the orbital coefficients, and with them the similarity $\psi$, depend on sectors B and C; and the density depends on sector B alone. The weights in (b) and (c) are inverse orbital-energy differences. On QH9, the HOMO and LUMO are about 0.25\,Ha apart and the deepest core orbital lies about 19\,Ha below the LUMO (Table~\ref{tab:datastats}), so an error coupling the two frontier orbitals changes the LUMO about 75 times more than an equally large error coupling it to the deepest core orbital (median over QH9; 5--95\% range 51--109). Item (c) is the first-order form of the Davis--Kahan theorem \citep{davis1970rotation}: the occupied subspace, and with it the density, is rotated only by sector B, with a strength set by the gap (Appendix~\ref{app:proofs}).
\end{remark}

\paragraph{Proof of Propositions~\ref{prop:sectors} and~\ref{prop:second}.} (a), (b) and (d) are Rayleigh--Schr\"odinger perturbation theory for the pencil $(\Hs+\dH,\Sm)$ after whitening \citep{kato1995perturbation,stewart1990perturbation}. For (c), $\delta\Pm=2\sum_{i\in\occ}(\delta\mathbf{c}_i\mathbf{c}_i^\top+\mathbf{c}_i\delta\mathbf{c}_i^\top)$; inserting (b) and pairing the terms $(i,j)$ and $(j,i)$ with $i,j\in\occ$ gives $\Delta_{ji}/(\varepsilon_i-\varepsilon_j)+\Delta_{ij}/(\varepsilon_j-\varepsilon_i)=0$ by symmetry of $\Dl$, which leaves
\[
\delta\Pm=2\sum_{i\in\occ}\sum_{a\in\vir}\frac{\Delta_{ai}}{\varepsilon^\star_i-\varepsilon^\star_a}\big(\mathbf{c}^\star_i\mathbf{c}^{\star\top}_a+\mathbf{c}^\star_a\mathbf{c}^{\star\top}_i\big)
\]
and \Eqref{eq:drho}. Part (c) does not need the non-degeneracy assumption. $\hat\Pm/2$ is the spectral projector of the pencil onto its $n_o$ lowest eigenvalues, which is analytic in $\Dl$ while $\|\Dl\|_2<g/2$ \citep{kato1995perturbation}; expanding its resolvent integral to first order gives entries $\Delta_{ai}/(\varepsilon^\star_i-\varepsilon^\star_a)$ between occupied and virtual orbitals and zero elsewhere, whether or not levels within $\occ$ or within $\vir$ coincide. The pairwise cancellation above is the same computation for non-degenerate levels. Beyond first order, the $\sin\Theta$ theorem of \citet{davis1970rotation} gives the same picture. In the reference orbital basis, the predicted orbitals are the eigenvectors of $\Cs{}^\top\Hh\Cs=\diag(\epsv^\star)+\Dl$, and the occupied reference subspace leaves a residual whose only nonzero block is $\Dl_{\vir\occ}$; if $\|\Dl\|_2<g/2$, the canonical angles $\Theta$ between the occupied subspaces of the reference and the prediction, which measure how far the occupied subspace, and with it $\hat\Pm$ and $\hat\rho$, has moved, therefore satisfy $\|\sin\Theta\|_F\le\|\Dl_{\occ\vir}\|_F/(g-2\|\Dl\|_2)$ \citep{stewart1990perturbation}. (e) is the variational principle for the Kohn--Sham energy with respect to occupied--virtual rotations.

\paragraph{Proof of Proposition~\ref{prop:blind}.} If $\Dl_{\occ\vir}=0$ then $\Cs{}^\top\Hh\Cs$ is block diagonal in $(\occ,\vir)$, so $\spn\{\Cs_{\occ}\}$ is an $\Sm$-invariant subspace of $\Hh$; if no eigenvalue crosses the gap it is the occupied subspace of $\Hh$, and $\hat\Pm=\Pm^\star$ because $\Pm$ depends only on the subspace. The dimension count is that of symmetric matrices on $\occ$ and on $\vir$.

\paragraph{Remark (directions versus norm).} Proposition~\ref{prop:blind} counts directions, not norm: at the QH9 median, $69\%$ of the parameters of $\Dl$ lie in the two blind sectors (Table~\ref{tab:datastats}), and this is the room that density matching leaves unpenalized. How much of an actual residual sits in that room is a separate, empirical question. For the physically realized Hamiltonian changes of Appendix~\ref{app:realresidual}, the blind sectors carry $15.6\%$ of $\|\Dl\|_F^2$ at a displacement of $0.02$\,\AA{} and $30.3\%$ at $0.30$\,\AA{}, almost all of the growth in the virtual--virtual block that contains the LUMO energy error. The blind share is therefore smaller in norm than in dimension, but it contains every orbital-energy error and grows under distribution shift.

\clearpage
\subsection{Properties of the cumulative LDOS loss}
Theorem~\ref{thm:props} collects the following results.

\begin{proposition}[Fermi-level identity; Theorem~\ref{thm:props}(i)]
\label{prop:identity}
Let $(\varepsilon_i,\psi_i)$ be the orbitals of $\Hs$ or of $\Hh$, and let the energy $E$ lie between their HOMO and LUMO energies at a distance of at least $t\sigma$ from both. Then, at every point $\mathbf{r}$, $|\Nre(\mathbf{r},E)-\rho(\mathbf{r})/2|\le\tfrac12\erfc(t/\sqrt2)\sum_i|\psi_i(\mathbf{r})|^2$: each orbital deviates by at most a fraction $\tfrac12\erfc(t/\sqrt2)$ of its density. At $E=\mu$ and for the reference, $t=g/(2\sigma)$, and the fraction is $3\times10^{-7}$ for $g=0.2$\,Ha and $\sigma=0.02$\,Ha.
\end{proposition}
\begin{proof}
By \Eqref{eq:density}, $\Nre(\mathbf{r},E)-\tfrac12\rho(\mathbf{r})=-\sum_{i\in\occ}|\psi_i(\mathbf{r})|^2\big[1-\Phi\big((E-\varepsilon_i)/\sigma\big)\big]+\sum_{a\in\vir}|\psi_a(\mathbf{r})|^2\,\Phi\big((E-\varepsilon_a)/\sigma\big)$. Every bracket and every $\Phi$ in this expression is at most $1-\Phi(t)=\tfrac12\erfc(t/\sqrt2)$, and the two sums have opposite signs, so the total is bounded by the larger of them.
\end{proof}
The identity holds at every energy inside the gap, not only at $\mu$. This matters for the training grid: with $K=64$ points on $[\mu-1.5,\mu+0.3]$\,Ha, $\mu$ is not a grid point but lies midway between the points $\mu\pm0.014$\,Ha. With $t=(g/2-0.014\,\mathrm{Ha})/\sigma$, both are density slices up to $2\times10^{-8}$ per orbital at the median QH9 gap and $1\times10^{-4}$ at its 5th percentile, so the loss contains density matching through these two slices (Table~\ref{tab:verify}). Orbitals far below the window contribute a constant step equal to their density, so nothing that density matching sees is lost; the other grid points add how that density is distributed in energy.

\begin{theorem}[No blind sector and identifiability; Theorem~\ref{thm:props}(ii, iii)]
\label{thm:complete}
Let the reference orbital energies be non-degenerate, and let $I\subseteq\mathbb{R}$ be an open energy interval, for example the loss window. Write $\varphi$ for the standard normal density, $\Phi_i(E):=\Phi((E-\varepsilon^\star_i)/\sigma)$ and $\varphi_i(E):=\varphi((E-\varepsilon^\star_i)/\sigma)$. Then
\begin{enumerate}[label=(\roman*)]
\item the first-order response $\Dl\mapsto\delta\Nre|_{\mathbb{R}^3\times I}$ is injective on symmetric $\Dl$: energy errors $d_i$ enter through the step positions with weight $|\psi_i(\mathbf{r})|^2\varphi_i(E)/\sigma$, and every off-diagonal $\Delta_{ij}$, including the occupied--occupied and virtual--virtual entries invisible to $\rho$, enters through the step heights with the $E$-profile $\Phi_i(E)-\Phi_j(E)$, a smoothed indicator of the interval between $\varepsilon^\star_i$ and $\varepsilon^\star_j$;
\item $\Nre[\Hh]=\Nre[\Hs]$ on $\mathbb{R}^3\times I$ if and only if $\Hh=\Hs$, and then $\hat\varepsilon_i=\varepsilon^\star_i$ and $\hat{\mathbf{c}}_i=\pm\mathbf{c}^\star_i$ for every $i$. By contrast, $\Lrho$ vanishes on the blind set of Proposition~\ref{prop:blind}, and the diagonal-only WALoss vanishes whenever $d_i=0$ for every $i$ with $w_i>0$.
\end{enumerate}
\end{theorem}
\begin{proof}
We use one fact about the energy profiles. \emph{(F) For distinct $\lambda_1,\dots,\lambda_m$, the $2m$ functions $\Phi((E-\lambda_k)/\sigma)$ and $\varphi((E-\lambda_k)/\sigma)$ are linearly independent on $I$.} All of them extend to entire functions of $E$, so a combination $\sum_k[\alpha_k\Phi((E-\lambda_k)/\sigma)+\beta_k\varphi((E-\lambda_k)/\sigma)]$ that vanishes on $I$ vanishes on $\mathbb{R}$. Differentiating it and multiplying by $\sqrt{2\pi}\,\sigma\,e^{E^2/2\sigma^2}$ gives $\sum_k\big[\alpha_k-\beta_k(E-\lambda_k)/\sigma\big]e^{-\lambda_k^2/2\sigma^2}e^{\lambda_kE/\sigma^2}=0$, and exponential polynomials with distinct exponents vanish only if every polynomial coefficient does, so $\alpha_k=\beta_k=0$ for all $k$.

(i) By Proposition~\ref{prop:sectors}(a,b), the first-order change of \Eqref{eq:cumldos} is
\[
\delta\Nre(\mathbf{r},E)=\sum_{i=1}^{n}\Big[2\psi_i(\mathbf{r})\,\delta\psi_i(\mathbf{r})\,\Phi_i(E)-\frac{d_i}{\sigma}\,\psi_i(\mathbf{r})^2\,\varphi_i(E)\Big],\qquad
\delta\psi_i=\sum_{j\ne i}\frac{\Delta_{ji}}{\varepsilon^\star_i-\varepsilon^\star_j}\,\psi_j\in\mathcal{U}.
\]
Collecting the terms $(i,j)$ and $(j,i)$ gives $2\Delta_{ij}\,\psi_i\psi_j\,[\Phi_i(E)-\Phi_j(E)]/(\varepsilon^\star_i-\varepsilon^\star_j)$, the step-height profile of the statement. If $\delta\Nre=0$ on $\mathbb{R}^3\times I$, then (F), applied at each $\mathbf{r}$, gives $\psi_i\,\delta\psi_i=0$ and $d_i\psi_i^2=0$ on $\mathbb{R}^3$ for every $i$. Since $\psi_i\ne0$, the product property of $\mathcal{U}$ gives $\delta\psi_i=0$, and $d_i=0$; linear independence of the $\psi_j$ then gives $\Delta_{ji}=0$ for all $j\ne i$. Hence $\Dl=\mathbf{0}$.

(ii) For a Hamiltonian $\mathbf{H}$ with eigenpairs $(\varepsilon_i,\mathbf{c}_i)$, group \Eqref{eq:cumldos} by distinct eigenvalues, $\Nre[\mathbf{H}](\mathbf{r},E)=\sum_\lambda w_\lambda(\mathbf{r})\,\Phi((E-\lambda)/\sigma)$ with $w_\lambda=\sum_{i:\,\varepsilon_i=\lambda}\psi_i^2$. If $\Nre[\Hh]=\Nre[\Hs]$ on $\mathbb{R}^3\times I$, (F) applied to the union of the two spectra gives $\hat w_\lambda=w^\star_\lambda$ on $\mathbb{R}^3$ for every $\lambda$. For an eigenvalue $\lambda$ of $\Hh$, $\hat w_\lambda$ is a sum of squares of nonzero functions in $\mathcal{U}$, so it is not identically zero and $\lambda$ is an eigenvalue of $\Hs$; the converse holds likewise. $\Hs$ has $n$ distinct eigenvalues and $\Hh$ has $n$ eigenvalues counted with multiplicity, so every eigenvalue of $\Hh$ is simple and $\hat\varepsilon_i=\varepsilon^\star_i$. Then $\hat\psi_i^2=\psi_i^2$, that is, $(\hat\psi_i-\psi_i)(\hat\psi_i+\psi_i)=0$ on $\mathbb{R}^3$, so $\hat\psi_i=\pm\psi_i$ and $\hat{\mathbf{c}}_i=\pm\mathbf{c}^\star_i$. Finally, \Eqref{eq:gep} gives $\mathbf{H}=\Sm\mathbf{C}\diag(\epsv)\mathbf{C}^\top\Sm$ for either Hamiltonian, and the signs cancel, so $\Hh=\Hs$; the converse is immediate. The statements about $\Lrho$ and the diagonal-only WALoss are Proposition~\ref{prop:blind} and Lemma~\ref{lem:whitening}.
\end{proof}
The theorem concerns the representation $\Nre$ over all of space; the loss is its sample estimate at 256 points and $K=64$ energies per molecule (Section~\ref{sec:impl}). It needs nothing of the basis beyond linear independence and the analyticity of Gaussian-type functions. We state it on all of space rather than on a finite point set because the natural finite-sample condition, linear independence of the products $\phi_\mu\phi_\nu$ on the points, fails in def2-SVP: on an atom with two $p$ shells of radial parts $R_1,R_2$, $p_{1x}p_{2y}=xyR_1R_2=p_{1y}p_{2x}$ identically. Because $I$ may be any interval, the theorem applies to the loss window, but orbital pairs far outside the window enter only through the Gaussian tails of their profiles, which is the energy-window weighting of Table~\ref{tab:coverage}. Table~\ref{tab:verify} checks both statements on QH9 molecules. On the level-0 quadrature grid, the first-order response is injective over the full energy range, and on the loss window it resolves every direction that involves an orbital inside the window, although some only weakly. The theorem does not carry over to the 256 points sampled per molecule: there, some directions on the loss window remain unresolved, mostly couplings between occupied orbitals and virtual orbitals above the window, whose energy profiles are nearly constant over the window, and less than 1\% of their weight lies on pairs of orbitals inside the window. The element-wise term supervises these directions, and a denser sample resolves them.

\begin{proposition}[Non-saturating restoring force; Theorem~\ref{thm:props}(iv)]
\label{prop:saturation}
Displace one orbital energy by $\delta$ with its shape fixed. Per sample point and up to the factor $|\psi_i(\mathbf{r})|^4$, the mismatch integrated over $E$ is, for a unit-height Gaussian peak,
$g(\delta)=2\sqrt\pi\,\sigma\big(1-e^{-\delta^2/4\sigma^2}\big)$, which saturates at $2\sqrt\pi\sigma$ with $g'(\delta)=(\sqrt\pi\,\delta/\sigma)\,e^{-\delta^2/4\sigma^2}\to0$; and, for the cumulative step,
$c(\delta)=\delta\,\erf\!\big(\tfrac{\delta}{2\sigma}\big)+\tfrac{2\sigma}{\sqrt\pi}\big(e^{-\delta^2/4\sigma^2}-1\big)$, with $c'(\delta)=\erf\!\big(\tfrac{\delta}{2\sigma}\big)\to\sgn\delta$. As $\sigma\to0$, $\Lcum$ becomes the Cram\'er--von Mises distance between the local spectral measures $\sum_i|\psi_i(\mathbf{r})|^2\delta_{\varepsilon_i}$ of $\Hh$ and $\Hs$.
\end{proposition}
\begin{proof}
For any square-integrable $f$, $\int(f(E-\delta)-f(E))^2dE=2\|f\|_2^2-2(f\star f)(\delta)$ with $f\star f$ the autocorrelation. For the unit-height Gaussian $f(E)=e^{-E^2/2\sigma^2}$, $\|f\|_2^2=\sigma\sqrt\pi$ and $(f\star f)(\delta)=\sigma\sqrt\pi\,e^{-\delta^2/4\sigma^2}$, giving $g(\delta)$. For the step, differentiate $c(\delta)=\int(\Phi((E-\delta)/\sigma)-\Phi(E/\sigma))^2dE$ under the integral: $c'(\delta)=\tfrac{2}{\sigma}\int(\Phi(E/\sigma)-\Phi((E-\delta)/\sigma))\varphi((E-\delta)/\sigma)\,dE=\erf(\delta/2\sigma)$, the last step being the Gaussian convolution identity $\int\Phi(x)\varphi(x-a)dx$ evaluated in closed form; integrating from $c(0)=0$ gives the stated $c(\delta)$. Both closed forms agree with numerical quadrature to machine precision (Table~\ref{tab:verify}). As $\sigma\to0$, $c(\delta)\to|\delta|$, the $W_1$ distance between two unit point masses, and for general spectra $\int(\Nre[\Hh]-\Nre[\Hs])^2dE$ is the Cram\'er--von Mises statistic between the two local spectral measures.
\end{proof}
A peak stops pulling once it no longer overlaps its target; a step does not, as long as the displaced level stays inside the energy window (Figure~\ref{fig:sector}b,c). At $\sigma=0.02$\,Ha and an energy error of 0.1\,Ha, the peak gradient has fallen to 1\% of its maximum, while the cumulative gradient is at 99.96\% of its limit. On a window $[a,b]$, the part of the integral for $c'(\delta)$ that the window omits is bounded by $\tfrac2\sigma\int\varphi((x-\delta)/\sigma)\,dx$ over $x\notin[a-\varepsilon^\star_i,\,b-\varepsilon^\star_i]$, so
\[
\Big|c_{[a,b]}'(\delta)-\erf\Big(\frac{\delta}{2\sigma}\Big)\Big|\le2\,\Phi\Big(\frac{a-\varepsilon^\star_i-\delta}{\sigma}\Big)+2\,\Phi\Big(\frac{\varepsilon^\star_i+\delta-b}{\sigma}\Big).
\]
The restoring force is therefore that of the proposition, up to an exponentially small error, while the displaced level is a few $\sigma$ inside the window, and it vanishes, like the peak's, once the level has left the window. The step is also wide in $E$, so the coarse training grid does not alias it: on the $K=64$ grid, whose spacing is $1.43\sigma$, the gradient stays within $0.01$ of $\erf(\delta/2\sigma)$ for $|\delta|\le0.1$\,Ha and every level at least $0.18$\,Ha inside the window (Table~\ref{tab:verify}).

\clearpage
\subsection{Error reallocation}
\begin{proposition}[Error reallocation]
\label{prop:realloc}
Let $\mathcal{M}\not\ni0$ be the set of errors a model can reach and $W$ a sector weighting with large entries on $d_i$ and $\Dl_{\occ\vir}$. If $\Dl_{\mathrm{iso}}$ minimizes $\|\Dl\|_F^2$ over $\mathcal{M}$ and $\Dl_w$ minimizes $\|\Dl\|_F^2+w\|W\!\circ\!\Dl\|_F^2$, then $\|\Dl_w\|_F\ge\|\Dl_{\mathrm{iso}}\|_F$ and $\|W\!\circ\!\Dl_w\|_F\le\|W\!\circ\!\Dl_{\mathrm{iso}}\|_F$, both strictly unless the two objectives share a minimizer.
\end{proposition}
\begin{proof}
$\Dl_w$ is feasible for the first problem, so $\|\Dl_{\mathrm{iso}}\|^2\le\|\Dl_w\|^2$; $\Dl_{\mathrm{iso}}$ is feasible for the second, so $\|\Dl_w\|^2+w\|W\!\circ\!\Dl_w\|^2\le\|\Dl_{\mathrm{iso}}\|^2+w\|W\!\circ\!\Dl_{\mathrm{iso}}\|^2$; combining the two gives both inequalities, and equality in either forces a common minimizer.
\end{proof}
Every eigen-derived metric is a function of the $W$-weighted sectors, while the complementary sectors $\Dl_{\occ\occ},\Dl_{\vir\vir}$ hold most of the degrees of freedom. A spectral loss can therefore lower every eigen-derived metric at the price of at most a small increase in the element-wise error. The two losses also push on $\Hh$ differently. With $f_E(x):=\Phi((E-x)/\sigma)$, the Daleckii--Krein formula for the derivative of a spectral function \citep{bhatia1997matrix} gives $\nabla_{\Hh}\Lcum=\Ch\,\mathbf{G}\,\Ch{}^\top$ with
\[
G_{ij}=\frac{2}{|\mathcal{R}|K}\sum_{\mathbf{r}\in\mathcal{R}}\sum_{k=1}^{K}\frac{\Nre[\Hh](\mathbf{r},E_k)-\Nre[\Hs](\mathbf{r},E_k)}{\mathrm{div}(\mathbf{r})^2}\,f^{[1]}_{ij}(E_k)\,\hat\psi_i(\mathbf{r})\hat\psi_j(\mathbf{r}),
\]
where $f^{[1]}_{ij}(E)=\big(f_E(\hat\varepsilon_i)-f_E(\hat\varepsilon_j)\big)/(\hat\varepsilon_i-\hat\varepsilon_j)$ for $i\ne j$ and $f^{[1]}_{ii}(E)=f_E'(\hat\varepsilon_i)$. The divided difference is bounded by $\min\{1/(\sqrt{2\pi}\sigma),\,1/|\hat\varepsilon_i-\hat\varepsilon_j|\}$ and is exponentially small unless a grid energy lies between $\hat\varepsilon_i$ and $\hat\varepsilon_j$ or within a few $\sigma$ of them. The gradient of $\Lcum$ is thus concentrated on close and frontier pairs of predicted orbitals inside the window, whereas the gradient of the mean-absolute element-wise loss, $\sgn(\dH)/n^2$, weights every entry of $\dH$ alike. Read in reverse, the proposition also explains why continuing to minimize the element-wise loss alone on a converged model can move error \emph{into} the sensitive sectors (Appendix~\ref{app:honly}).

%% file: refs.bib
@inproceedings{yu2023qhnet,
  title={Efficient and Equivariant Graph Networks for Predicting Quantum {H}amiltonian},
  author={Yu, Haiyang and Xu, Zhao and Qian, Xiaofeng and Qian, Xiaoning and Ji, Shuiwang},
  booktitle={International Conference on Machine Learning (ICML)},
  year={2023}
}

@inproceedings{yu2023qh9,
  title={{QH9}: A Quantum {H}amiltonian Prediction Benchmark for {QM9} Molecules},
  author={Yu, Haiyang and Liu, Meng and Luo, Youzhi and Strasser, Alex and Qian, Xiaofeng and Qian, Xiaoning and Ji, Shuiwang},
  booktitle={Advances in Neural Information Processing Systems (NeurIPS), Datasets and Benchmarks Track},
  year={2023}
}

@inproceedings{li2025wanet,
  title={Enhancing the Scalability and Applicability of {K}ohn-{S}ham {H}amiltonians for Molecular Systems},
  author={Li, Yunyang and Xia, Zaishuo and Huang, Lin and Wei, Xinran and Harshe, Samuel and Yang, Han and Luo, Erpai and Wang, Zun and Zhang, Jia and Liu, Chang and Shao, Bin and Gerstein, Mark},
  booktitle={International Conference on Learning Representations (ICLR)},
  year={2025}
}

@inproceedings{luo2025sphnet,
  title={Efficient and Scalable Density Functional Theory {H}amiltonian Prediction through Adaptive Sparsity},
  author={Luo, Erpai and Wei, Xinran and Huang, Lin and Li, Yunyang and Yang, Han and Xia, Zaishuo and Wang, Zun and Liu, Chang and Shao, Bin and Zhang, Jia},
  booktitle={International Conference on Machine Learning (ICML)},
  year={2025}
}

@inproceedings{unke2021phisnet,
  title={{SE(3)}-equivariant prediction of molecular wavefunctions and electronic densities},
  author={Unke, Oliver T. and Bogojeski, Mihail and Gastegger, Michael and Geiger, Mario and Smidt, Tess and M{\"u}ller, Klaus-Robert},
  booktitle={Advances in Neural Information Processing Systems (NeurIPS)},
  year={2021}
}

@article{schutt2019schnorb,
  title={Unifying machine learning and quantum chemistry with a deep neural network for molecular wavefunctions},
  author={Sch{\"u}tt, Kristof T. and Gastegger, Michael and Tkatchenko, Alexandre and M{\"u}ller, Klaus-Robert and Maurer, Reinhard J.},
  journal={Nature Communications},
  volume={10},
  pages={5024},
  year={2019}
}

@article{li2022deeph,
  title={Deep-learning density functional theory {H}amiltonian for efficient ab initio electronic-structure calculation},
  author={Li, He and Wang, Zun and Zou, Nianlong and Ye, Meng and Xu, Runzhang and Gong, Xiaoxun and Duan, Wenhui and Xu, Yong},
  journal={Nature Computational Science},
  volume={2},
  pages={367--377},
  year={2022}
}

@inproceedings{zhang2024selfconsistency,
  title={Self-Consistency Training for Density-Functional-Theory {H}amiltonian Prediction},
  author={Zhang, He and Liu, Chang and Wang, Zun and Wei, Xinran and Liu, Siyuan and Zheng, Nanning and Shao, Bin and Liu, Tie-Yan},
  booktitle={International Conference on Machine Learning (ICML)},
  year={2024}
}

@article{gu2022eigen,
  title={Neural network representation of electronic structure from ab initio molecular dynamics},
  author={Gu, Qiangqiang and Zhang, Linfeng and Feng, Ji},
  journal={Science Bulletin},
  volume={67},
  number={1},
  pages={29--37},
  year={2022}
}

@article{ellis2021ldos,
  title={Accelerating finite-temperature {K}ohn-{S}ham density functional theory with deep neural networks},
  author={Ellis, J. Austin and Fiedler, Lenz and Popoola, Gabriel A. and Modine, Normand A. and Stephens, John A. and Thompson, Aidan P. and Cangi, Attila and Rajamanickam, Sivasankaran},
  journal={Physical Review B},
  volume={104},
  pages={035120},
  year={2021}
}

@article{jorgensen2022deepdft,
  title={Equivariant graph neural networks for fast electron density estimation of molecules, liquids, and solids},
  author={J{\o}rgensen, Peter Bj{\o}rn and Bhowmik, Arghya},
  journal={npj Computational Materials},
  volume={8},
  pages={183},
  year={2022}
}

@article{koker2024charge3net,
  title={Higher-order equivariant neural networks for charge density prediction in materials},
  author={Koker, Teddy and Quigley, Keegan and Taw, Eric and Tibbetts, Kevin and Li, Lin},
  journal={npj Computational Materials},
  volume={10},
  pages={161},
  year={2024}
}

@article{ushenin2025lagnet,
  title={{LAGNet}: better electron density prediction for {LCAO}-based data and drug-like substances},
  author={Ushenin, Konstantin and Khrabrov, Kuzma and Tsypin, Artem and Ber, Anton and Rumiantsev, Egor and Kadurin, Artur},
  journal={Journal of Cheminformatics},
  volume={17},
  pages={65},
  year={2025}
}

@article{hohenberg1964,
  title={Inhomogeneous electron gas},
  author={Hohenberg, Pierre and Kohn, Walter},
  journal={Physical Review},
  volume={136},
  number={3B},
  pages={B864--B871},
  year={1964}
}

@article{kohn1965,
  title={Self-consistent equations including exchange and correlation effects},
  author={Kohn, Walter and Sham, Lu Jeu},
  journal={Physical Review},
  volume={140},
  number={4A},
  pages={A1133--A1138},
  year={1965}
}

@article{kohn1999nobel,
  title={Nobel Lecture: Electronic structure of matter---wave functions and density functionals},
  author={Kohn, Walter},
  journal={Reviews of Modern Physics},
  volume={71},
  number={5},
  pages={1253--1266},
  year={1999}
}

@article{tersoff1985stm,
  title={Theory of the scanning tunneling microscope},
  author={Tersoff, Jerry and Hamann, Donald R.},
  journal={Physical Review B},
  volume={31},
  number={2},
  pages={805--813},
  year={1985}
}

@article{selloni1985stm,
  title={Voltage-dependent scanning-tunneling microscopy of a crystal surface: Graphite},
  author={Selloni, Annabella and Carnevali, Paolo and Tosatti, Erio and Chen, C. D.},
  journal={Physical Review B},
  volume={31},
  number={4},
  pages={2602--2605},
  year={1985}
}

@book{stewart1990perturbation,
  title={Matrix Perturbation Theory},
  author={Stewart, Gilbert W. and Sun, Ji-guang},
  publisher={Academic Press},
  year={1990}
}

@article{pulay1980diis,
  title={Convergence acceleration of iterative sequences. {T}he case of {SCF} iteration},
  author={Pulay, P{\'e}ter},
  journal={Chemical Physics Letters},
  volume={73},
  number={2},
  pages={393--398},
  year={1980}
}

@article{sun2018pyscf,
  title={{PySCF}: the {P}ython-based simulations of chemistry framework},
  author={Sun, Qiming and Berkelbach, Timothy C. and Blunt, Nick S. and Booth, George H. and Guo, Sheng and Li, Zhendong and Liu, Junzi and McClain, James D. and Sayfutyarova, Elvira R. and Sharma, Sandeep and Wouters, Sebastian and Chan, Garnet Kin-Lic},
  journal={WIREs Computational Molecular Science},
  volume={8},
  number={1},
  pages={e1340},
  year={2018}
}

@article{ramakrishnan2014qm9,
  title={Quantum chemistry structures and properties of 134 kilo molecules},
  author={Ramakrishnan, Raghunathan and Dral, Pavlo O. and Rupp, Matthias and von Lilienfeld, O. Anatole},
  journal={Scientific Data},
  volume={1},
  pages={140022},
  year={2014}
}

@book{kato1995perturbation,
  author    = {Kato, Tosio},
  title     = {Perturbation Theory for Linear Operators},
  series    = {Classics in Mathematics},
  publisher = {Springer},
  address   = {Berlin},
  year      = {1995}
}

@book{bhatia1997matrix,
  author    = {Bhatia, Rajendra},
  title     = {Matrix Analysis},
  series    = {Graduate Texts in Mathematics},
  volume    = {169},
  publisher = {Springer},
  address   = {New York},
  year      = {1997}
}

@article{davis1970rotation,
  author  = {Davis, Chandler and Kahan, W. M.},
  title   = {The rotation of eigenvectors by a perturbation. {III}},
  journal = {SIAM Journal on Numerical Analysis},
  volume  = {7},
  number  = {1},
  pages   = {1--46},
  year    = {1970}
}

@book{szabo1996modern,
  author    = {Szabo, Attila and Ostlund, Neil S.},
  title     = {Modern Quantum Chemistry: Introduction to Advanced Electronic Structure Theory},
  publisher = {Dover},
  address   = {Mineola, NY},
  year      = {1996}
}

@inproceedings{kaniselvan2025helm,
  title={Learning from the Electronic Structure of Molecules across the Periodic Table},
  author={Kaniselvan, Manasa and Miller, Benjamin Kurt and Gao, Meng and Nam, Juno and Levine, Daniel S.},
  booktitle={International Conference on Learning Representations (ICLR)},
  year={2026}
}

@inproceedings{arjovsky2017wgan,
  title={{W}asserstein Generative Adversarial Networks},
  author={Arjovsky, Martin and Chintala, Soumith and Bottou, L{\'e}on},
  booktitle={International Conference on Machine Learning (ICML)},
  year={2017}
}
